**Chapter in the book "Electromagnetic Metamaterials"**

**Effective Medium Theory of Electromagnetic and Quantum Metamaterials**

Mário G. Silveirinha

*University of Lisbon, Instituto Superior Técnico-Instituto de Telecomunicações*
*Torre Norte, Av. Rovisco Pais 1, 1049-001 Lisboa, Portugal*
*mario.silveirinha@co.it.pt*

Here, we present an overview of recent developments in the characterization of electromagnetic and quantum metamaterials using effective medium methods. It is highlighted that both electromagnetic and electronic systems can be homogenized in a unified manner based on the introduction of an effective Hamiltonian operator that describes the time evolution of the macroscopic initial states as well as the stationary states of the relevant system. Furthermore, it is shown that in some circumstances quadratic forms of the fields, such as the energy, can be exactly determined using the effective medium theory.

## 1. Introduction

The wave propagation in periodic or in random systems can be rather challenging to study as an exact treatment of problem is generally unfeasible and a numerical analysis with no approximations requires large scale time-consuming computations. In addition, brute force numerical calculations provide very limited insights of the physical mechanisms that determine the wave propagation. Fortunately, it is often possible to describe "low-energy" wave phenomena by resorting to an effective medium description that regards the structure of interest as a continuum.

For example, natural materials are formed by a collection of atoms or molecules arranged – in case of crystalline structures – in a periodic lattice. Even within a classical framework wherein the atoms are regarded as electric or magnetic dipoles, the propagation of light in a natural material can be immensely complex: each atom or molecule scatters and absorbs light, and the propagation problem must be solved self-consistently. Auspiciously, provided the light wavelength is much larger than the lattice constant, i.e., larger than the characteristic spacing between the atoms, the complex light-matter interactions can be homogenized and the material can be regarded as continuum described by a certain effective permittivity and permeability [1-4]. Indeed, natural materials may often be regarded as continuous media without any granularity for light wavelengths as short as some tens of nanometers.

As a second example, consider the propagation of electron waves under the influence of an ionic lattice in the context of one-body Schrödinger equation, e.g. the propagation of electrons in a semiconductor material. The electron wave function is scattered by the electric potential created by the ionic lattice, and hence the time evolution of a given initial electronic state is challenging to characterize. Fortunately, for low energy phenomena the ionic lattice can be effectively homogenized in such a manner that its effect on the wave propagation can be described by an effective mass [5]. In other words, the medium may be regarded as a continuum provided the electron mass is suitably redefined to take into account the influence of the scattering centers.

The objective of this Chapter is to present a unified overview of the recent developments in the research of the wave propagation in complex media using effective medium methods, for both light and electron waves. The contents of this Chapter are largely based on the ideas originally introduced in Refs. [6, 7], and which were further developed in subsequent works.

## 2. Microscopic theory

We consider a generic physical system whose dynamics is characterized by a one-body Schrödinger-type equation of the form:

$$\hat{H}\psi = i\hbar \frac{\partial}{\partial t}\psi \ . \tag{1}$$

Here, $\hbar$ is the reduced Planck constant, $\hat{H}$ is the Hamiltonian operator that determines the time evolution of the system and $\psi$ is the wave function that describes the state of the system. In general $\psi$ is a multi-component vector (a spinor). Evidently, this type of formulation is suitable to characterize the propagation of electron waves in a bulk semiconductor or in a semiconductor superlattice. Interestingly, the propagation of light can also be described using a similar formulation. Indeed, the Maxwell's equations can be written in a compact form as [7]

$$\begin{pmatrix} 0 & i\nabla \times \mathbf{1}_{3\times 3} \\ -i\nabla \times \mathbf{1}_{3\times 3} & 0 \end{pmatrix} \cdot \mathbf{f} = i\frac{\partial \mathbf{g}}{\partial t} \ , \tag{2}$$

where $\mathbf{f} = \begin{pmatrix} \mathbf{e} & \mathbf{h} \end{pmatrix}^T$ is a six-element vector with components determined by the microscopic electric and magnetic fields and $\mathbf{g} = \begin{pmatrix} \mathbf{d} & \mathbf{b} \end{pmatrix}^T$ is a six-element vector with components determined by the electric displacement and the magnetic induction fields. In electromagnetic metamaterials the $\mathbf{g}$ and $\mathbf{f}$ fields are related by a space-dependent material matrix $\mathbf{M} = \mathbf{M}(\mathbf{r})$ through the constitutive relation $\mathbf{g} = \mathbf{M} \cdot \mathbf{f}$. In conventional isotropic media the material matrix is simply:

$$\mathbf{M} = \begin{pmatrix} \varepsilon \mathbf{1}_{3\times3} & 0 \\ 0 & \mu \mathbf{1}_{3\times3} \end{pmatrix}. \tag{3}$$

Hence, by introducing the operator $\hat{H}$ given by:

$$\hat{H} = \hbar \begin{pmatrix} 0 & i\nabla \times \mathbf{1}_{3\times3} \\ -i\nabla \times \mathbf{1}_{3\times3} & 0 \end{pmatrix} \cdot \mathbf{M}^{-1}. \tag{4}$$

and identifying the state vector with the $\mathbf{g}$ field, $\psi = \mathbf{g}$, the Maxwell's equations can be expressed as in Eq. (1). It should be noted that in the electromagnetic case $\hat{H}$ is unrelated to the energy of the system, and should be simply regarded as an operator that describes the time evolution of the classical electromagnetic field. Moreover, in the previous discussion it is implicit that the relevant materials are nondispersive, i.e., the permittivity $\varepsilon$ and the permeability $\mu$ are frequency independent. Yet, the previous ideas can be generalized to dispersive media. Indeed, for dispersive materials the multiplication operator $\mathbf{M}^{-1}$ should be replaced by a suitable time convolution operator. In that case, the action of $\hat{H}$ on $\psi$ at a generic time instant $t$ depends not only on $\psi$ at the same time instant, but also on the values of the state vector in the past $t' < t$. Thus, for dispersive media Eq. (1) should be understood as a generalized Schrödinger-type equation.

In the following subsections, we further elaborate on some physical systems to which the theory applies.

## 2.1. *Electromagnetic metamaterials*

The first case of interest corresponds to that of electromagnetic metamaterials, i.e., mesoscopic structures formed by dielectric or metallic inclusions arranged in a periodic lattice [8]. The electromagnetic response of metamaterials is mainly determined by the geometry of its constituents, rather than directly by the chemical composition. The light propagation in these structures is described by the Maxwell's equations, which, as previously discussed, can be recast in the Schrödinger form with the time evolution operator given by Eq. (4). An illustrative metamaterial unit cell is represented in Figure 1a, and corresponds to a spherical dielectric inclusion embedded in a metallic region (the host material). The three-dimensional metamaterial is formed by the periodic repetition of the unit cell. The problem of homogenization of metamaterials, the extraction of effective medium parameters, and the limitations inherent to an effective medium description have been extensively discussed in the literature [9-26].

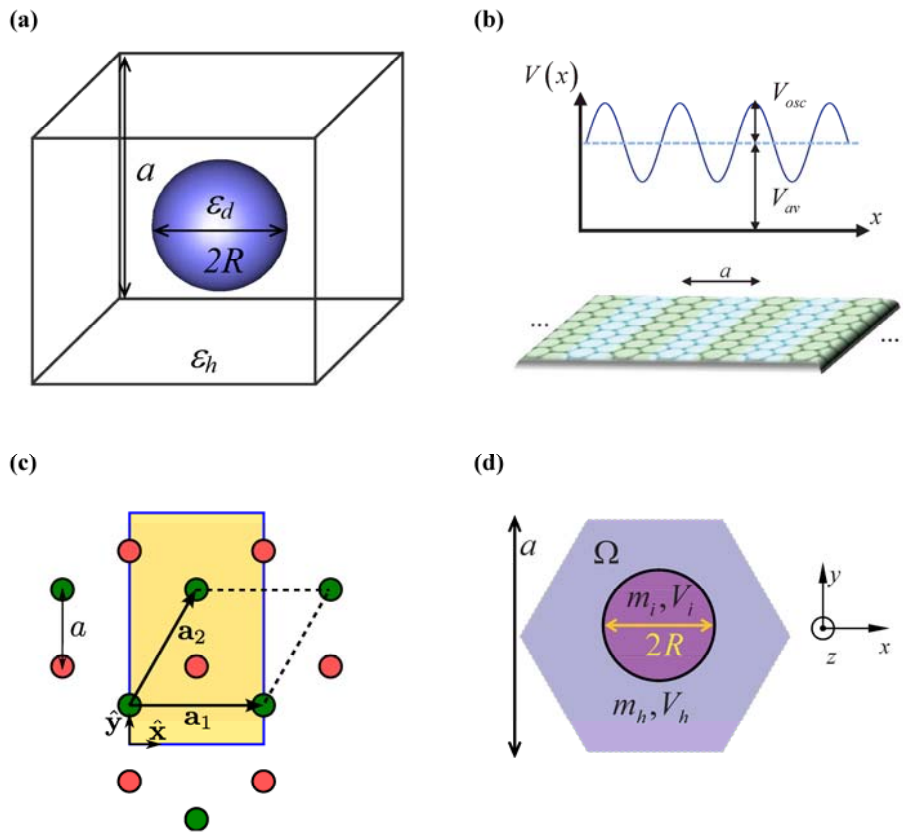

Fig. 1. Sketch of the geometries of several metamaterial and semiconductor superlattices. (a) Unit cell of an electromagnetic metamaterial formed by a cubic array of spherical dielectric inclusions embedded in a host medium. (b) Graphene superlattice created by a periodic electrostatic potential. (c) Nanopatterned 2D-electron gas. (d) Unit cell of a semiconductor superlattice formed by two different materials.

## 2.2. *Semiconductor superlattices*

In a groundbreaking study, Esaki and Tsu suggested in 1969 that new quantum effects can be observed at an intermediate physical scale if a mono-crystalline semiconductor is either periodically doped or if the composition of a semiconductor alloy is periodically varied [27]. Such structures are known as semiconductor superlattices and can be regarded as the precursors and the semiconductor counterparts of modern electromagnetic metamaterials. The original proposal of Esaki and Tsu has set the stage for the engineering of the electron transport [27-29], and lead to the development of novel electronic materials with ultrahigh mobilities wherein the electrons can experience a near zero effective mass, and other breakthroughs [30-32].

The time evolution of the electron wave function in a semiconductor superlattice may be characterized by a Hamiltonian of the form:

$$\hat{H} = -\frac{\hbar^2}{2} \nabla \cdot \left(\frac{1}{m}\nabla\right) + V \,, \tag{5}$$

where $V$ is a static electric potential and $m$ is the electron (effective) mass in the pertinent semiconductor. Usually, $V$ and $m$ are periodic functions of the spatial coordinates. For example, the above Hamiltonian may describe the physics of a two-dimensional (2D) electron gas (e.g. a semiconductor quantum well) modulated by an electrostatic potential created by nanopatterned scattering centers (Figure 1c) [33]. A different possibility is to periodically change the material composition (Figure 1d). In this case, the superlattice has the hexagonal symmetry and is formed by two different semiconductors.

### 2.3. *Graphene superlattices*

Graphene is a one-atom thick material discovered at the turn of the 21st century [34-35]. Remarkably, graphene has a relativistic-type electronic spectrum such that the relation between energy and momentum is linear [36]. This unusual property enables mimicking quantum relativistic effects in a condensed-matter platform, and creates many exciting opportunities in nanoelectronics [36-38]. Interestingly, it is possible to tailor the electronic transport in graphene using the superlattice concept [39-45]. This can be done with patterned gates that impress a space dependent electric potential on a graphene sheet (Figure 1b). Other solutions take advantage of the electric potential induced by crystalline substrates such as boron nitride, or create a desired electric potential through the controlled deposition of adatoms. The low-energy electronics in graphene near the $K$ point is described by a two-dimensional massless Dirac Hamiltonian [35]

$$\hat{H} = -i\hbar v_F \left(\boldsymbol{\sigma} \cdot \nabla\right) + V \,, \tag{6}$$

where $v_F \sim 10^6 \, m/s$ is the Fermi velocity, $\boldsymbol{\sigma} = \left( \boldsymbol{\sigma}_x, \boldsymbol{\sigma}_y \right)$, $\boldsymbol{\sigma}_x, \boldsymbol{\sigma}_y$ are the Pauli matrices, and $V$ is the periodic electrostatic potential. The wave function $\psi$ in graphene is a pseudo-spinor, and hence has two components [35]. From a physical point of view, each component of the pseudo-spinor is associated with a specific trigonal sublattice of graphene.

## 3. Effective medium theory

The goal of an effective medium theory is to provide an approximate and simplified description of the wave propagation in some complex system. Due to the spatially inhomogeneous nature of the structure, the wave function (in case of electronic systems) or the electromagnetic fields (in case of light waves) can vary wildly in the characteristic length scale determined by the "granularities". In this work, we restrict our attention to periodic structures, and hence the characteristic length scale is defined by the lattice period. Usually one is interested in "low-energy" phenomena for which the wave packet envelope varies slowly in space. An effective medium theory aims to describe the dynamics of the wave packet envelope. These ideas will be made more precise in the following subsections.

### 3.1. *Spatial averaging and the envelope function*

The envelope function is intuitively the slowly varying in space part of the state vector $\psi$. It is defined here as:

$$\Psi\left(\mathbf{r}, t\right) \equiv \left\{ \psi\left(\mathbf{r}, t\right) \right\}_{\mathrm{av}}, \tag{7}$$

where $\left\{ \ \right\}_{\mathrm{av}}$ is a linear operator that performs a spatial averaging. The averaging operator is completely determined by the response to plane waves, determined by the function $F\left(\mathbf{k}\right)$ such that

$$\left\{ e^{i\mathbf{k} \cdot \mathbf{r}} \right\}_{\mathrm{av}} = F\left(\mathbf{k}\right) e^{i\mathbf{k} \cdot \mathbf{r}}. \tag{8}$$

Thus, the action of the averaging operator on a generic plane wave with wave vector $\mathbf{k}$ yields another plane wave with the same wave vector, but with a different amplitude determined by $F\left(\mathbf{k}\right)$. Because of the linearity of the operator $\left\{ \ \right\}_{\mathrm{av}}$, its action on a generic function is determined by Fourier theory and is given by a spatial convolution. For example, it is possible to write the envelope function as:

$$\Psi\left(\mathbf{r}, t\right) = \int d^N \mathbf{r}' \, f\left(\mathbf{r}'\right) \psi\left(\mathbf{r} - \mathbf{r}', t\right), \tag{9}$$

where $N$ is the space dimension (e.g., $N=3$ for a three-dimensional metamaterial, and $N=2$ for graphene). The weight function $f$ is the inverse Fourier transform of $F$ so that:

$$f(\mathbf{r}) = \frac{1}{(2\pi)^N} \int d^N\mathbf{k}\, F(\mathbf{k}) e^{i\mathbf{k}\cdot\mathbf{r}} \,. \tag{10}$$

Related ideas have been developed by Russakov in the context of macroscopic electromagnetism [46]. In this work, it is assumed that the averaging operator is corresponds to an ideal low pass spatial filter such that:

$$F(\mathbf{k}) = \begin{cases} 1, & \mathbf{k} \in \text{B.Z.} \\ 0, & \text{otherwise} \end{cases} \,. \tag{11}$$

In the vast majority of the cases of interest the set B.Z. stands for the first Brillouin zone of the periodic lattice, but sometimes other choices can be relevant. Unless something different is explicitly stated, it will always be assumed that B.Z. is the first Brillouin zone.

With these definitions, the envelope function $\Psi(\mathbf{r},t)$ has no relevant spatial fluctuations on the scale of a unit cell, i.e., the microscopic fluctuations are filtered out by the averaging operator. Hence, we will also refer to $\Psi(\mathbf{r},t)$ as the macroscopic state vector. In general, we say that a given state vector $\psi$ is macroscopic when it stays invariant under the operation of spatial averaging:

$$\psi(\mathbf{r}) = \left\{\psi(\mathbf{r})\right\}_{\text{av}}, \quad \text{(macroscopic state vector)}. \tag{12}$$

Importantly, a macroscopic state cannot be more localized in space than the characteristic period of the material.

### 3.2. *The effective Hamiltonian*

The effective Hamiltonian is the operator that describes the time evolution of the envelope function. Specifically, suppose that the initial state vector is macroscopic, so that $\psi_{t=0} = \Psi_{t=0}$. In general, the time evolution of an initial macroscopic state does not yield a macroscopic state at a later time instant, i.e., $\psi(\mathbf{r},t) \neq \Psi(\mathbf{r},t)$ for $t>0$. We define the effective Hamiltonian $\hat{H}_{ef}$ in such a manner that $\Psi(\mathbf{r},t)$ calculated using $\hat{H}_{ef}$ is coincident with the spatially-averaged microscopic state vector $\left\{\psi(\mathbf{r},t)\right\}_{\text{av}}$, being $\psi(\mathbf{r},t)$ determined by the microscopic Hamiltonian $\hat{H}$ [7, 47]. These ideas are illustrated in the diagram of Figure 2.

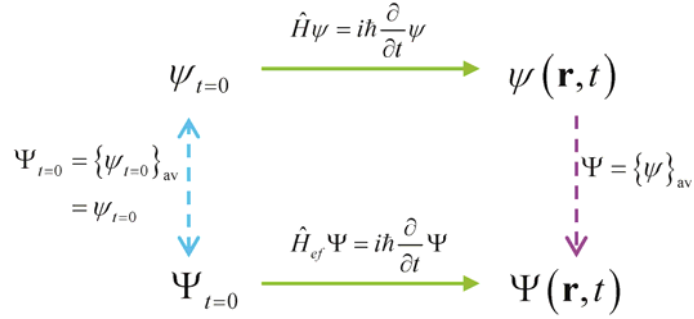

Fig. 2. Schematic relation between the time evolutions determined by the macroscopic and microscopic Hamiltonians: for an initial macroscopic state the effective medium formulation ensures that $\Psi = \{\psi\}_{av}$ for $t > 0$. Reprinted with permission [47].

The time evolution of the macroscopic state vector is determined by a generalized Schrödinger equation:

$$\left( \hat{H}_{ef} \Psi \right) (\mathbf{r}, t) = i\hbar \frac{\partial}{\partial t} \Psi (\mathbf{r}, t) . \tag{13}$$

From the definition of the effective Hamiltonian it is clear that it must ensure that:

$$\left\{ \hat{H} \psi \right\}_{av} = \hat{H}_{ef} \Psi . \tag{14}$$

It was shown in Ref. [7] that the action of the effective Hamiltonian on the wave function can be written as a convolution in space and in time:

$$\left( \hat{H}_{ef} \Psi \right) (\mathbf{r}, t) = \int d^N \mathbf{r}' \int_0^t dt' \, \mathbf{h}_{ef} \left( \mathbf{r} - \mathbf{r}', t - t' \right) \cdot \Psi (\mathbf{r}', t') . \tag{15}$$

In general, the kernel $\mathbf{h}_{ef} (\mathbf{r}, t)$ is represented by a square matrix $\left[ \mathbf{h}_{\sigma, \sigma'} \right]$ because $\Psi$ is typically a multi-component vector. In the electronic case the dimension of $\mathbf{h}$ is determined by the number $S$ of spin or pseudospin degrees of freedom ($\sigma = 1, ..., S$), whereas in the light case the dimension of $\mathbf{h}_{ef}$ is $S$=6. Equation (15) implies that the effective Hamiltonian depends on the past history ($0 < t' < t$) of the macroscopic state vector, rather than just on the instantaneous value of $\Psi$.

It is convenient to introduce the Fourier transform of $\mathbf{h}_{ef} (\mathbf{r}, t)$ defined as:

$$\mathcal{H}_{ef} (\mathbf{k}, \omega) = \int d^N \mathbf{r} \int_0^{+\infty} dt \, \mathbf{h}_{ef} (\mathbf{r}, t) e^{i\omega t} e^{-i\mathbf{k} \cdot \mathbf{r}} . \tag{16}$$

The Fourier transform is bilateral in space and unilateral in time. The unilateral Fourier transform in time can also be regarded as a Laplace transform. In the Fourier domain, the action of the effective Hamiltonian reduces to a simple multiplication:

$$\left(\hat{H}_{ef}\Psi\right)\left(\mathbf{k},\omega\right) = \mathcal{H}_{ef}\left(\mathbf{k},\omega\right)\cdot\Psi\left(\mathbf{k},\omega\right). \tag{17}$$

In the above, $\Psi\left(\mathbf{k},\omega\right)$ is the Fourier transform of the macroscopic state vector,

$$\Psi\left(\mathbf{k},\omega\right) = \int d^N\mathbf{r}\int_0^{+\infty} dt\,\Psi\left(\mathbf{r},t\right)e^{i\omega t}e^{-i\mathbf{k}\cdot\mathbf{r}}, \tag{18}$$

and $\left(\hat{H}_{ef}\Psi\right)\left(\mathbf{k},\omega\right)$ is defined similarly. The convergence of the Fourier transforms is ensured for $\text{Im}\{\omega\} > 0$. The function $\mathcal{H}_{ef}\left(\mathbf{k},\omega\right)$ completely determines the effective Hamiltonian. Because of the properties of the spatial averaging operator, it is possible to enforce that:

$$\mathcal{H}_{ef}\left(\mathbf{k},\omega\right) = 0, \qquad \text{when } \mathbf{k} \notin \text{B.Z.}. \tag{19}$$

This property ensures that the effective Hamiltonian is a smoothened version of the microscopic Hamiltonian. In the next subsections, it is explained how $\mathcal{H}_{ef}\left(\mathbf{k},\omega\right)$ can be calculated for $\mathbf{k} \in \text{B.Z.}$.

### 3.3. *Calculation of $\mathcal{H}_{ef}\left(\mathbf{k},\omega\right)$ with a time domain approach*

Let us consider an initial macroscopic state of the form $\psi_{t=0} \sim e^{i\mathbf{k}\cdot\mathbf{r}}\mathbf{u}_l$ with $\mathbf{k} \in \text{B.Z.}$. Here, $\left(\mathbf{u}_l\right)$ represents a basis of unit vectors that generates the $S$-dimensional vector space wherein $\psi$ is defined. Because of the periodicity of the system, the microscopic time evolution of this initial state yields a state vector $\psi\left(\mathbf{r},t\right)$ with the Bloch property. Specifically, $\psi\left(\mathbf{r},t\right)e^{-i\mathbf{k}\cdot\mathbf{r}}$ is a periodic function in space for any fixed $t$. For the same reason, $\hat{H}\psi$ has also the Bloch property. Importantly, the operation of spatial averaging only retains spatial harmonics with wave vector inside the B.Z., and hence it follows that the dependence of $\{\psi\}_{\text{av}}$ and $\{\hat{H}\psi\}_{\text{av}}$ on the spatial coordinates is of the form $e^{i\mathbf{k}\cdot\mathbf{r}}$ for any time instant. In other words, within the effective medium approach the time evolution of a plane wave-type initial state yields another plane wave-type state. Moreover, it is possible to write:

$$\{\psi\}_{\text{av}}\left(\mathbf{r},t\right) = \psi_{\text{av}}\left(t\right)e^{i\mathbf{k}\cdot\mathbf{r}}, \tag{20a}$$

$$\{\hat{H}\psi\}_{\text{av}}\left(\mathbf{r},t\right) = \left(\hat{H}\psi\right)_{\text{av}}\left(t\right)e^{i\mathbf{k}\cdot\mathbf{r}}, \tag{20b}$$

with

$$\psi_{\text{av}}\left(t\right) = \frac{1}{V_{cell}} \int_{\Omega} d^N \mathbf{r} \, \psi\left(\mathbf{r}, t\right) e^{-i\mathbf{k} \cdot \mathbf{r}}, \tag{21a}$$

$$\left(\hat{H}\psi\right)_{\text{av}}\left(t\right) = \frac{1}{V_{cell}} \int_{\Omega} d^N \mathbf{r} \, \hat{H}\psi\left(\mathbf{r}, t\right) e^{-i\mathbf{k} \cdot \mathbf{r}}, \tag{21b}$$

where $\Omega$ represents the unit cell and $V_{cell}$ is the respective volume.

Taking now into account that $\Psi = \left\{\psi\right\}_{\text{av}}$ and $\hat{H}_{ef}\Psi = \left\{\hat{H}\psi\right\}_{\text{av}}$, and substituting Eq. (20) into Eq. (15), it is seen that:

$$\left(\hat{H}\psi\right)_{\text{av}}\left(\omega\right) = \mathcal{H}_{ef}\left(\mathbf{k}, \omega\right) \cdot \psi_{\text{av}}\left(\omega\right). \tag{22}$$

In the above, $\psi_{\text{av}}\left(\omega\right)$ and $\left(\hat{H}\psi\right)_{\text{av}}\left(\omega\right)$ stand for the unilateral Fourier (Laplace) transforms of the functions in Eq. (21). Hence, if we denote $\psi^{(l)}$, $l=1,\ldots,S$ as the microscopic state vector determined by the time evolution of the initial state $\psi_{t=0}^{(l)} = i/\hbar \, e^{i\mathbf{k}\cdot\mathbf{r}}\mathbf{u}_l$ (the proportionality constant was fixed as $i/\hbar$ for convenience), it follows from the previous analysis that the effective Hamiltonian is given by:

$$\mathcal{H}_{ef}\left(\mathbf{k}, \omega\right) = \left[\left(\hat{H}\psi^{(1)}\right)_{\text{av}} \quad \ldots \quad \left(\hat{H}\psi^{(S)}\right)_{\text{av}}\right] \cdot \left[\psi_{\text{av}}^{(1)} \quad \ldots \quad \psi_{\text{av}}^{(S)}\right]^{-1}. \tag{23}$$

Thus, $\mathcal{H}_{ef}\left(\mathbf{k}, \omega\right)$ can be written as the product of two matrices, whose columns are determined by the vectors $\psi_{\text{av}}^{(l)}\left(\omega\right)$ and $\left(\hat{H}\psi^{(l)}\right)_{\text{av}}\left(\omega\right)$.

In summary, for a given $\mathbf{k} \in$ B.Z. the effective Hamiltonian can be found by solving $S$ microscopic time evolution problems associated with initial states of the form $\psi_{t=0}^{(l)} = i/\hbar \, e^{i\mathbf{k}\cdot\mathbf{r}}\mathbf{u}_l$. The effective Hamiltonian is written in terms of the Fourier transforms in time of the functions (21).

### 3.4.   *Calculation of $\mathcal{H}_{ef}\left(\mathbf{k}, \omega\right)$ with a frequency domain approach*

The effective Hamiltonian may also be determined based on frequency domain calculations. To prove this we note that $\psi_{\text{av}}\left(\omega\right)$ and $\left(\hat{H}\psi\right)_{\text{av}}\left(\omega\right)$ can be written explicitly as:

$$\psi_{\text{av}}\left(\omega\right) = \frac{1}{V_{cell}} \int_{\Omega} d^N \mathbf{r} \, \psi\left(\mathbf{r}, \omega\right) e^{-i\mathbf{k} \cdot \mathbf{r}}, \tag{24a}$$

$$\left(\hat{H}\psi\right)_{\text{av}}\left(\omega\right) = \frac{1}{V_{cell}} \int_{\Omega} d^N \mathbf{r} \, \hat{H}\psi\left(\mathbf{r}, \omega\right) e^{-i\mathbf{k} \cdot \mathbf{r}}, \tag{24b}$$

where $\psi(\mathbf{r},\omega)$ is the unilateral Fourier transform of $\psi(\mathbf{r},t)$. Applying the unilateral Fourier (Laplace) transform to both members of the microscopic Schrödinger equation (1) and using the property $\partial_t \psi(\mathbf{r},t) \leftrightarrow -i\omega\psi(\mathbf{r},\omega) - \psi_{t=0}(\mathbf{r})$, it follows that:

$$\left[\hat{H} - \hbar\omega\right] \cdot \psi(\mathbf{r},\omega) = -i\hbar\psi_{t=0}(\mathbf{r}). \tag{25}$$

Hence, $\psi^{(l)}(\mathbf{r},\omega)$ can be directly found by solving the above equation for $-i\hbar\psi_{t=0}^{(l)} = e^{i\mathbf{k}\cdot\mathbf{r}}\mathbf{u}_l$, with $l=1,\ldots,S$. Once $\psi^{(l)}(\mathbf{r},\omega)$ is known one can determine $\psi_{\mathrm{av}}^{(l)}$ and $\left(\hat{H}\psi^{(l)}\right)_{\mathrm{av}}$ using Eq. (24), and finally obtain the effective Hamiltonian from Eq. (23).

It is interesting to note that for $-i\hbar\psi_{t=0}^{(l)} = e^{i\mathbf{k}\cdot\mathbf{r}}\mathbf{u}_l$ Eq. (25) implies that $\left(\hat{H}\psi^{(l)}\right)_{\mathrm{av}} - \hbar\omega\psi_{\mathrm{av}}^{(l)} = \mathbf{u}_l$. Substituting this result into Eq. (23) one may also write the effective Hamiltonian as:

$$\mathcal{H}_{ef}(\mathbf{k},\omega) = \hbar\omega + \left[\psi_{\mathrm{av}}^{(1)} \quad \ldots \quad \psi_{\mathrm{av}}^{(s)}\right]^{-1}. \tag{26}$$

### 3.5. *The electromagnetic case*

In the case of light waves, using the time evolution operator (4) and $\psi = \mathbf{g} = (\mathbf{d} \quad \mathbf{b})^T$, it is possible to rewrite Eq. (25) as:

$$\begin{pmatrix} 0 & i\nabla\times\mathbf{1}_{3\times3} \\ -i\nabla\times\mathbf{1}_{3\times3} & 0 \end{pmatrix} \cdot \mathbf{f}(\mathbf{r}) - \omega\mathbf{g}(\mathbf{r}) = -i\mathbf{g}_{t=0}(\mathbf{r}), \tag{27}$$

with $\mathbf{f} = (\mathbf{e} \quad \mathbf{h})^T$. If we decompose the six-vector $\mathbf{g}_{t=0}$ as $\mathbf{g}_{t=0} = (\mathbf{j}_e \quad \mathbf{j}_m)^T$, the above system can be spelled out as:

$$\begin{aligned} \nabla\times\mathbf{e} &= +i\omega\mathbf{b} - \mathbf{j}_m \\ \nabla\times\mathbf{h} &= -i\omega\mathbf{d} + \mathbf{j}_e \end{aligned}. \tag{28}$$

These correspond to the standard microscopic Maxwell's equations in the frequency domain with fictitious electric-type and magnetic-type sources, $\mathbf{j}_e$ and $\mathbf{j}_m$, respectively. Clearly, in the homogenization problem the sources have a plane-wave spatial dependence $e^{i\mathbf{k}\cdot\mathbf{r}}$. Thus, the effective response of a composite medium can be calculated by exciting the medium with fictitious macroscopic sources. This idea is the essence of the "source-driven" homogenization method originally introduced in Ref. [6]. Next, we prove that the effective response obtained using the approach of section 3.4 is coincident with what is obtained using the theory of Refs. [6, 17-21]. Related effective medium formalisms have also been presented in Refs. [22-26].

To this end, define $\mathbf{G}_{av} = \begin{pmatrix} \mathbf{D}_{av} & \mathbf{B}_{av} \end{pmatrix}^T$ and $\mathbf{F}_{av} = \begin{pmatrix} \mathbf{E}_{av} & \mathbf{H}_{av} \end{pmatrix}^T$ such that (compare with Eq. 24a):

$$\mathbf{G}_{av} = \frac{1}{V_{cell}} \int_{\Omega} d^N \mathbf{r} \ \mathbf{g}(\mathbf{r}) e^{-i\mathbf{k} \cdot \mathbf{r}}, \tag{29a}$$

$$\mathbf{F}_{av} = \frac{1}{V_{cell}} \int_{\Omega} d^N \mathbf{r} \ \mathbf{f}(\mathbf{r}) e^{-i\mathbf{k} \cdot \mathbf{r}}. \tag{29b}$$

Furthermore, let us introduce the effective material matrix $\mathbf{M}_{ef}(\mathbf{k}, \omega)$ such that for arbitrary macroscopic sources $\mathbf{j}_e$ and $\mathbf{j}_m$ one has:

$$\mathbf{G}_{av} = \mathbf{M}_{ef}(\mathbf{k}, \omega) \cdot \mathbf{F}_{av}. \tag{30}$$

Then, it can be shown from Eqs. (4) and (22) that:

$$\mathcal{H}_{ef}(\mathbf{k}, \omega) = \hbar \begin{pmatrix} 0 & -\mathbf{k} \times \mathbf{1}_{3\times3} \\ \mathbf{k} \times \mathbf{1}_{3\times3} & 0 \end{pmatrix} \cdot \mathbf{M}_{ef}^{-1}(\mathbf{k}, \omega). \tag{31}$$

This proves that the effective Hamiltonian can be written in terms of the effective material matrix $\mathbf{M}_{ef}(\mathbf{k}, \omega)$ calculated with the source-driven homogenization [6]. In particular, the time evolution of the macroscopic electromagnetic fields can be determined from Eq. (13), which is equivalent to the macroscopic Maxwell's equations:

$$\begin{pmatrix} 0 & i\nabla \times \mathbf{1}_{3\times3} \\ -i\nabla \times \mathbf{1}_{3\times3} & 0 \end{pmatrix} \cdot \mathbf{F}(\mathbf{r}, t) = i \frac{\partial}{\partial t} \mathbf{G}(\mathbf{r}, t), \tag{32}$$

being $\mathbf{G} = \{\mathbf{g}\}_{av}$ and $\mathbf{F} = \{\mathbf{f}\}_{av}$ the macroscopic electromagnetic fields. Consistent with the conventional theory of spatially dispersive materials [48], the fields $\mathbf{G}$ and $\mathbf{F}$ are related by a space-time convolution whose kernel is determined by the inverse Fourier transform of $\mathbf{M}_{ef}(\mathbf{k}, \omega)$. It can be shown that for reciprocal structures the effective material matrix satisfies:

$$\mathbf{M}_{ef}(\mathbf{k}, \omega) = \mathbf{U} \cdot \mathbf{M}_{ef}^T(-\mathbf{k}, \omega) \cdot \mathbf{U}, \quad \text{with} \quad \mathbf{U} = \begin{pmatrix} \mathbf{1}_{3\times3} & 0 \\ 0 & -\mathbf{1}_{3\times3} \end{pmatrix}. \tag{33}$$

In most electromagnetic metamaterials, the inclusions are either dielectric or metallic particles, and thus do not have an intrinsic magnetic response ($\mu = \mu_0$). In this case, it is evident that independent of the excitation one has $\mathbf{B}_{av} = \mu_0 \mathbf{H}_{av}$. This result together with the reciprocity constraint (33) imply that for metamaterials formed by non-magnetic particles the material matrix is of the form:

$$\mathbf{M}_{ef}\left(\mathbf{k},\omega\right)=\begin{pmatrix} \overline{\varepsilon}_{ef}\left(\mathbf{k},\omega\right) & 0 \\ 0 & \mu_0\mathbf{1}_{3\times3} \end{pmatrix}. \tag{34}$$

Thus, the effective response of metal-dielectric metamaterials is completely characterized by a nonlocal dielectric function $\overline{\varepsilon}_{ef}\left(\mathbf{k},\omega\right)$, consistent with Ref. [6].

In summary, it was proven that the time evolution of macroscopic electromagnetic field states characterized by a certain $\mathbf{g}_{t=0}$ is rigorously described by the operator $\mathcal{H}_{ef}\left(\mathbf{k},\omega\right)$ given by Eq. (31). The effective Hamiltonian is written in terms of an effective material matrix $\mathbf{M}_{ef}\left(\mathbf{k},\omega\right)$, which is exactly coincident with that originally introduced in Ref. [6], based on the idea of source-driven homogenization.

### 3.6. *Stationary states*

A key property of the effective Hamiltonian is that its energy spectrum coincides with that of the microscopic Hamiltonian [7]. The energy spectrum of the macroscopic Hamiltonian is determined by the nontrivial solutions of the stationary Schrödinger equation

$$\left[\left.\mathcal{H}_{ef}\left(\mathbf{k},\omega\right)\right|_{\omega=E/\hbar}-E\right]\cdot\Psi=0\,, \tag{35}$$

where $E$ stands for the energy of a certain stationary state. Likewise, in the electromagnetic case the photonic band structure calculated with the effective Hamiltonian is coincident with exact band structure obtained using a microscopic theory [6].

The enunciated result can be understood noting that in a time evolution problem (with no source excitation) the state vector can be written as a superposition of eigenmodes. The eigenmodes have a time variation of the form $e^{-i\omega_n t}$, being $\omega_n=E_n/\hbar$ the relevant eigenfrequencies. Importantly, since the macroscopic and microscopic state vectors are related by the spatial-averaging operation ($\Psi=\{\psi\}_{av}$), both $\Psi$ and $\psi$ the same-type of time oscillations. In other words, the averaging affects only the space coordinates, while the time coordinate is not averaged in any manner. As a consequence, the spectrum of the microscopic and macroscopic Hamiltonians must be the same [Eq. (35)].

Strictly speaking, it is possible that some special microscopic states are not predicted by the effective medium Hamiltonian. We refer to such states as "dark states" [7]. A dark state corresponds to a microscopic stationary state $\psi$ that has a zero projection into the subspace of macroscopic states, i.e., a state such that $\Psi = \{\psi\}_{av} = 0$. Dark states appear only in degenerate singular cases, and for very specific forms of the microscopic Hamiltonian. Thus, typically the band structures of the microscopic and effective Hamiltonians are indeed the same.

## 4.   Applications

Next, we present several examples that illustrate the application of the ideas developed in section 3 to both electromagnetic metamaterials and quantum structures.

### 4.1.   *Homogenization of electromagnetic metamaterials*

The interest in modern electromagnetic metamaterials was sparked by a seminal study by J. Pendry, who showed that a composite material with a simultaneously negative permittivity and permeability can make a perfect lens [49]. Among many other proposals, it was suggested that such a doubly negative metamaterial can be realized based on dielectric spherical particles embedded in a metallic host material [20, 50-51] (Figure 1a). The following discussion is focused on the metamaterial homogenization.

As mentioned in section 3.5, a metamaterial formed by dielectrics and metals is completely characterized by a nonlocal dielectric function of the form $\overline{\varepsilon}_{ef}(\mathbf{k}, \omega)$. Such a description is rather powerful but still quite complex. Indeed, both the frequency and the wave vector are independent parameters, and thus the nonlocal dielectric function depends on four independent variables. It is desirable to further simplify the theory. This can be done based on the hypothesis that the material response is local. Specifically, if one assumes that the homogenized medium can be characterized by a local permittivity, $\overline{\varepsilon}_L(\omega)$, and by a local permeability, $\overline{\mu}_L(\omega)$, then it can be shown that the local parameters are linked to the nonlocal dielectric function as [6, 21]:

$$\frac{1}{\varepsilon_0}\overline{\varepsilon}_{ef}(\mathbf{k}, \omega) = \overline{\varepsilon}_L(\omega) + \frac{c\mathbf{k}}{\omega} \times \left(\overline{\mu}_L^{-1}(\omega) - \mathbf{1}\right) \times \frac{c\mathbf{k}}{\omega}. \tag{36}$$

Note that the local parameters are independent of the wave vector. The above relation implies that the local permittivity satisfies:

$$\overline{\varepsilon}_L(\omega) = \frac{1}{\varepsilon_0}\overline{\varepsilon}_{ef}(\mathbf{k} = 0, \omega). \tag{37}$$

On the other hand, the local permeability can be written in terms of the second order derivatives of the nonlocal dielectric function with respect to the wave vector. For example, assuming that $\overline{\mu}_L(\omega)$ is a diagonal tensor it is easy to check that:

$$\mu_{L,zz}(\omega) = \frac{1}{1 - \left(\dfrac{\omega}{c}\right)^2 \dfrac{1}{2\varepsilon_0} \dfrac{\partial^2 \varepsilon_{ef,yy}}{\partial k_x^2}\bigg|_{\mathbf{k}=0}} \quad . \tag{38}$$

The formulas for the other components of the permeability can be obtained by considering permutations of the indices $x$, $y$, and $z$. In practice, $\overline{\varepsilon}_{ef}(\mathbf{k},\omega)$ is calculated using computational methods. Several solutions have been reported in the literature: an integral equation based approach [6], a finite-difference frequency domain method [19], and a time domain scheme [20]. The derivatives with respect to the wave vector are calculated using finite differences [6, 19, 20].

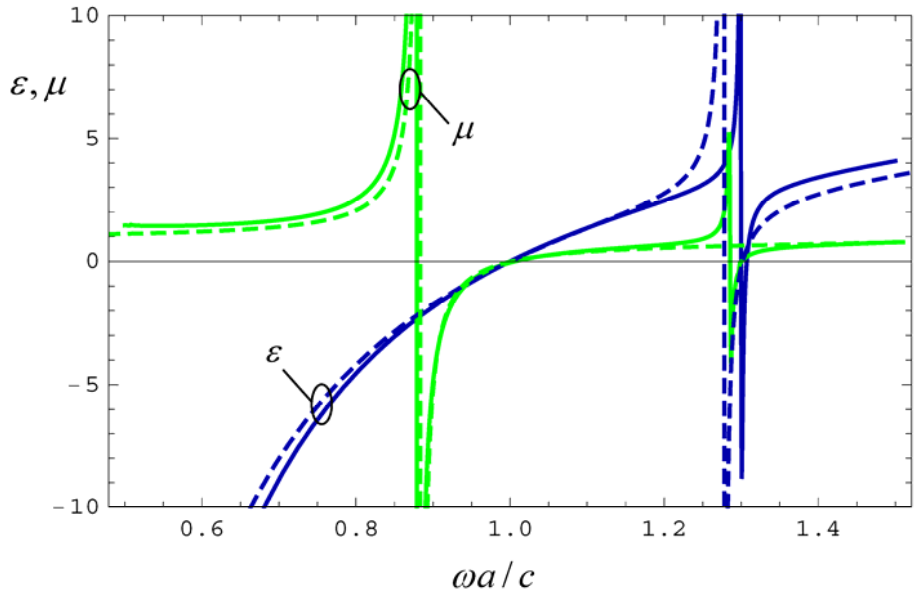

Fig. 3. Local effective permittivity $\varepsilon$ and local effective permeability $\mu$ as a function of frequency for a simple cubic array of dielectric spheres with $\varepsilon_d = 73.1$ and $R = 0.4a$ embedded in a metallic host. Solid lines: Effective medium theory; Dashed lines: Lewin's formulas [52]. Adapted from Ref. [20].

Figure 3 depicts the extracted local effective parameters for a metamaterial with unit cell as in Figure 1a [20]. For long wavelengths the metamaterial has an isotropic response, and hence both the permittivity and permeability are scalars.

The radius of the dielectric spheres is $R = 0.4a$, being $a$ the lattice constant. The spheres have permittivity $\varepsilon_d = 73.1$ and are embedded in a metallic material with a permittivity modeled by a Drude-type dispersion, $\varepsilon_h = \varepsilon_\infty \left( 1 - \omega_p^2 / \omega \left( \omega + i\Gamma \right) \right)$. In the simulations, it was assumed that the normalized plasma frequency is $\omega_p a / c = 1.0$, that $\varepsilon_\infty = 3.6$, and for simplicity the metallic loss was neglected $\Gamma = 0$. The parameters of the metamaterial were tuned to ensure that $\varepsilon_L \approx \mu_L$ (matched index material) over a broad range of frequencies close to $\omega = \omega_p$.

Figure 3 also shows the local effective parameters predicted by a mixing formula proposed by Lewin [52]. As seen, there is an overall excellent agreement between the full wave homogenization results and Lewin's formula, especially near $\omega = \omega_p$. Interestingly, in this design the conditions $\varepsilon = \mu = -1$ are observed simultaneously, and thus this metamaterial can mimic to some extent a Veselago-Pendry's lens, as further discussed in Ref. [51]. Many other examples of the application of the effective medium theory can be found in Refs. [6, 19, 20].

## 4.2.  *Zincblende semiconductors*

The electronic structure of semiconductors is typically studied using perturbation schemes, usually known as $k \cdot p$ methods [53-62]. The $k \cdot p$ theory is rooted on the knowledge of the electronic band structure at highly-symmetric points of the Brillouin zone determined by some eigenstates $u_{n0}$. These eigenstates are used as a basis to construct the wave function for a generic wave vector. Hence, the electron wave function is usually described by a multi-component vector, whose elements are the coefficients of the expansion of the wave function in the basis $u_{n0}$. The $k \cdot p$ theory can be applied to bulk materials as well as to semiconductor heterostructures.

In contrast, G. Bastard developed the concept of the envelope-function approximation in the analysis of semiconductor heterostructures during the 1980's [55, 61-62]. The idea of an envelope-function is closely related to the pseudopotential method used in condensed matter physics [5]. Interestingly, there is a profound connection between the effective medium methods developed in section 3 and Bastard's theory [47, 63]. Specifically, it was rigorously shown from "first principles" that when the effective medium approach is applied to bulk semiconductors with a zincblende structure – using as a starting point Kane's eight-band $k \cdot p$ theory [56] – the corresponding effective Hamiltonian is characterized by an energy dependent effective mass and an effective potential [47]. Moreover, the obtained effective Hamiltonian is equivalent in the long-wavelength limit to that used in Bastard's theory [55]. Thus, the formalism of section 3 recovers Bastard's theory and shows that the corresponding effective Hamiltonian describes rigorously the time evolution of the macroscopic electronic states in bulk semiconductors. Moreover, it puts into perspective that the envelope function approximation used in semiconductor physics is intrinsically related to the effective medium methods used in the context of macroscopic electrodynamics [47, 59].

The zincblende lattice corresponds to a face-centered cubic lattice with two atoms per unit cell, and is characteristic of binary III-V compounds such as GaAs, GaSb, InSb, and II-VI compounds such as HgTe and CdTe [53, 55]. The electronic structure of these materials is formed by a conduction band, a light-hole band, a heavy-hole band, and a split-off band [53, 57]. The wave function associated with the conduction band has the same symmetry as a monopole (*S*-type symmetry), while the wave functions associated with the remaining (valence) bands have the same symmetry as a dipole (*P*-type symmetry). The split-off band usually lies well below the remaining bands, and often its effects on the wave propagation can be neglected. The zincblende compounds can be characterized by the effective Hamiltonian [47]:

$$\mathcal{H}_{ef}\left(\mathbf{k}, E\right) = E_c + \frac{\hbar^2}{2m_0} k^2 \left[1 + \frac{\varepsilon_P}{3}\left(\frac{-2}{\tilde{E}_v} + \frac{-1}{\tilde{E}_v - \Delta}\right)\right]. \tag{39}$$

Here, $E = \hbar\omega$ is the electron energy, $\tilde{E}_v = E_v - E + \hbar^2 k^2 / \left(2m_0\right)$, $m_0$ is the free electron rest mass, $\Delta$ is the spin-orbit split-off energy, $\varepsilon_P = 2P^2 m_0 / \hbar^2$ is Kane's energy, and $E_c$ and $E_v$ determine the energy levels at the edges of the conduction and light-hole bands, respectively. The split-off energy $\Delta$ determines the energy offset between the split-off band and the other two valence bands.

For long wavelengths, the term $\hbar^2 k^2 / \left(2m_0\right)$ in the definition of $\tilde{E}_v$ can be neglected. Within this approximation, the effective Hamiltonian can be written as:

$$\hat{H}_{ef}\left(E\right) = -\frac{\hbar^2}{2}\nabla \cdot \left(\frac{1}{m_{ef}}\nabla\right) + V_{ef}, \tag{40}$$

being the effective potential given by

$$V_{ef}\left(E\right) = E_c, \tag{41}$$

and the dispersive effective mass $m_{ef} = m_{ef}\left(E\right)$ defined as

$$\frac{1}{m_{ef}} = \frac{1}{m_0} + v_P^2\left(\frac{2}{E - E_v} + \frac{1}{E - E_v + \Delta}\right), \tag{42}$$

where $v_P = \sqrt{\varepsilon_P/(3m_0)}$ is Kane's velocity. Thus, the electron wave propagation in the bulk semiconductor compound is completely characterized by an energy-dependent effective mass and by an effective potential. The effective mass formula (42) is well known in the context of Bastard's envelope function approximation [55, p. 88]. It is important to highlight that the effective parameter "dispersive mass" is distinct from the usual effective mass $m^* = \hbar^2 \left[ \partial^2 E / \partial k^2 \right]^{-1}$ obtained from the curvature of the energy diagram.

In case of narrow gap semiconductors, $\Delta$ is typically a few times larger than the band gap energy $\left| E_g \right| = \left| E_c - E_v \right|$. In these conditions, for energies in the band gap or in vicinity of the band gap only the first term in brackets in Eq. (42) is important. This yields the linear dispersive mass approximation:

$$m_{ef} \approx \frac{E - E_v}{2v_P^2}. \tag{43}$$

There is an interesting analogy between the propagation of electron waves in a zincblende semiconductor compound and the propagation of light waves in an isotropic material. Indeed, by comparing the stationary Schrödinger equation associated with the Hamiltonian (40) with the Helmholtz equation, $\nabla \cdot \left[ \mu^{-1} \nabla E_z \right] + \left( \omega^2 / c^2 \right) \varepsilon E_z = 0$, that describes the propagation of transverse electric (TE) electromagnetic waves (with $\mathbf{E} = E_z \hat{\mathbf{z}}$ and $\partial / \partial z = 0$) in a material characterized by the parameters $\varepsilon$ and $\mu$, it is possible to establish the following correspondence [47, 63, 64]:

$$\begin{aligned} E - V_{ef} &\leftrightarrow \varepsilon \\ m_{ef} &\leftrightarrow \mu \end{aligned}. \tag{44}$$

Thus $E - V_{ef}$ may be regarded as the semiconductor dual of the electric permittivity, whereas $m_{ef}$ may be regarded as the semiconductor dual of the magnetic permeability. This analogy can be useful to establish parallelisms between phenomena in electromagnetic metamaterials and in semiconductor superlattices. For example, in Ref. [63] it was shown how such a correspondence can be used to design novel semiconductor materials with extreme anisotropy, such that the effective mass is zero along some preferred direction of motion and infinite for perpendicular directions. Furthermore, based on these ideas it was theoretically suggested that the perfect lens concept and a light tunneling phenomenon in metamaterials have semiconductor analogues [64, 65].

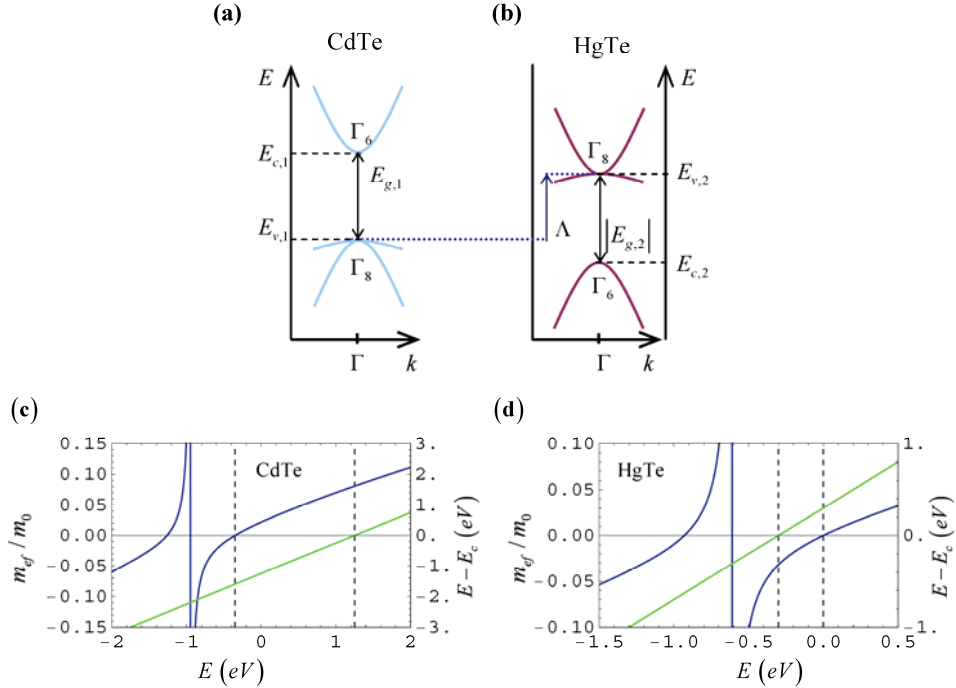

Fig. 4. (a) and (b) Sketch of the electronic band diagrams of CdTe and HgTe. (c) and (d) Effective parameters $m_{cf}$ and $E - V_{cf}$ as a function of the normalized energy for CdTe with a regular band structure and for HgTe with an inverted band structure. The energy of the valence band edge of HgTe is arbitrarily taken equal to zero and the valence band offset is $\Lambda = 0.35\,eV$. Blue (dark gray) lines: $m_{cf}/m_0$; Green (light gray) lines: $E - E_c$ in $[eV]$. The dashed vertical gridlines indicate the edges of the light-hole valence and conduction bands, and delimit the relevant bandgaps. Adapted from Refs. [47, 63].

It is interesting to further discuss the properties of the effective parameters of mercury-cadmium-telluride (HgCdTe) compounds. The electronic band structures of cadmium-telluride (CdTe) and of mercury-telluride (HgTe) (a group II-VI degenerate semiconductor) are sketched in Figures 4a and 4b, respectively. Following the usual convention, the conduction band with *S*-type symmetry is denoted by $\Gamma_6$, whereas the light-hole and heavy-hole valence bands with *P*-type symmetry are denoted by $\Gamma_8$. The heavy-hole band is the nearly flat valence band. As seen in Figure 4a, the binary compound CdTe has a regular band structure with the conduction band lying above the valence bands. Interestingly, even though unusual, it is possible that the order of the conduction band (with *S*-type symmetry) and of the valence bands (with *P*-type symmetry) is interchanged. An example of a material with this remarkable property is HgTe, which, as illustrated in Figure 4b, has an "inverted" band structure. Therefore, HgTe has a negative band gap energy ( $E_g = E_c - E_v < 0$ ) [47, 63, 64].

Figures 4c and 4d show the effective parameters $m_{ef}$ and $E - V_{ef}$ as a function of the electron energy for CdTe and HgTe, respectively. The stationary energy states occur at the energy intervals for which $m_{ef}$ and $E - V_{ef}$ have the same sign, and the band gaps occur in the range wherein $m_{ef}$ and $E - V_{ef}$ have different signs. The band gap that separates the conduction band from the light-hole valence band is delimited by the vertical dashed gridlines in Figure 4.

The plots in Figure 4 show that in the band gap a semiconductor with a regular band structure ( $E_g > 0$ , e.g., CdTe) behaves from the point of view of the wave propagation as a material with $\varepsilon < 0$ and $\mu > 0$ (*epsilon* negative –ENG– material), whereas a semiconductor with an inverted band structure ( $E_g < 0$ , e.g., HgTe) behaves as a material with $\varepsilon > 0$ and $\mu < 0$ (*mu* negative –MNG– material). Due to these remarkable properties the electron wave propagation in HgTe-CdTe superlattices exhibits remarkable analogies with the propagation of light in ENG-MNG metamaterials [63, 64, 65]. In general, it is estimated that the ternary semiconductor alloy Hg$_{1-x}$Cd$_x$Te exhibits a regular band structure for $x > 0.17$ and an inverted band structure for $x < 0.17$ , being $0 \leq x \leq 1$ the mole fraction of cadmium.

### 4.3. *Homogenization of semiconductor superlattices*

The effective Hamiltonian (40) enables a macroscopic description of the electron wave propagation in a bulk semiconductor, in the same manner as the permittivity and the permeability characterize the light propagation in a natural material. Similar to electromagnetic metamaterials, it is possible to tailor the transport properties of electrons by mixing different semiconductors in a periodic structure. To illustrate this, we consider the superlattice formed by two material phases arranged in a hexagonal lattice (Figure 1d). It is assumed that the circular scattering centers (the inclusions) have dispersive mass and effective potential $m_i, V_i$ , whereas the background material is characterized by the parameters $m_h, V_h$ . Furthermore, it is supposed that the electrons can move only in the *xoy* plane so that the structure is intrinsically two-dimensional. For example, the structure may correspond to a quantum well. In what follows, we assume that the host region is the ternary alloy Hg$_{0.75}$Cd$_{0.25}$Te and that the scattering centers are made of HgTe. Moreover, for simplicity the effective parameters $m, V$ are taken to be the same as in a bulk semiconductor.

The effective Hamiltonian of the superlattice can be found using the general approach outlined in section 3.4. In particular, to obtain $\mathcal{H}_{ef}(\mathbf{k}, E)$ one needs to solve Eq. (25) with respect to $\psi$ with $\hat{H}$ given by Eq. (5) where $m = m(\mathbf{r}, E)$ and $V = V(\mathbf{r})$ are periodic functions of the spatial coordinates. Because $\hat{H}$ is a scalar operator ($S=1$) one needs to solve a single microscopic problem with $\psi_{t=0} \sim e^{i\mathbf{k}\cdot\mathbf{r}}$. The effective Hamiltonian is given by $\mathcal{H}_{ef}(\mathbf{k}, E) = (\hat{H}\psi)_{\mathrm{av}} / \psi_{\mathrm{av}}$ with $(\hat{H}\psi)_{\mathrm{av}}$ and $\psi_{\mathrm{av}}$ defined as in Eq. (24). A finite difference frequency domain discretization of Eq. (25) is reported in Ref. [66].

It is desirable to further simplify the effective medium description. For low energy phenomena, this can be done relying on a Taylor expansion of the effective Hamiltonian with respect to the wave vector [47]:

$$\mathcal{H}_{ef}(\mathbf{k}, E) \approx V_{ef}(E) + \frac{\hbar^2}{2}\mathbf{k} \cdot \bar{M}_{ef}^{-1} \cdot \mathbf{k}, \qquad (45)$$

where $V_{ef}(E) = \mathcal{H}_{ef}(\mathbf{k} = 0, E)$ is the energy dependent effective potential of the superlattice and $\bar{M}_{ef}$ is the energy dependent effective mass tensor of the superlattice, being its inverse determined by

$$\bar{M}_{ef}^{-1} = \frac{1}{\hbar^2}\left[\frac{\partial^2 \mathcal{H}_{ef}}{\partial k_i \partial k_j}\bigg|_{\mathbf{k}=0}\right]. \qquad (46)$$

Note that similar to the effective permeability in electromagnetic metamaterials [Eq. (38)], the superlattice effective mass is written in terms of the second order derivatives of the effective Hamiltonian with respect to the wave vector.

For the particular superlattice under study, it is clear that by symmetry the effective mass tensor is scalar. Hence, within the validity of Eq. (45) the effective Hamiltonian in the space domain is such that:

$$\hat{H}_{ef}(\mathbf{r}, E) = -\frac{\hbar^2}{2}\nabla \cdot \left[\frac{1}{M_{ef}(E)}\nabla\right] + V_{ef}(E). \qquad (47)$$

Therefore, the electron wave propagation in the superlattice is completely characterized by the effective parameters $V_{ef}(E)$ and $M_{ef}(E)$. The order of the operators in the term $\nabla \cdot \left[M_{ef}^{-1}\nabla\right]$ is consistent with the generalized Ben Daniel-Duke boundary conditions [53, 55, 57]. Indeed, Eq. (47) implies that $\Psi$ and $M_{ef}^{-1}(\partial/\partial n)\Psi$ are continuous at an abrupt interface between two materials, which ensures the continuity of the normal component of the probability current at the interface.

Interestingly, it is possible to write approximate analytical formulas for $V_{ef}(E)$ and $M_{ef}(E)$. Indeed, by exploiting the correspondences $E - V \leftrightarrow \varepsilon$ and $m \leftrightarrow \mu$, and the analogies between electromagnetic metamaterials and semiconductor superlattices one finds that [67]:

$$V_{ef} = V_h \left(1 - f_V\right) + V_i f_V, \tag{48}$$

$$M_{ef} = m_h \frac{\left(1 - f_V\right) m_h + \left(1 + f_V\right) m_i}{\left(1 + f_V\right) m_h + \left(1 - f_V\right) m_i}, \tag{49}$$

where $f_V$ is the volume fraction of the HgTe inclusions. The mixing formula for the effective mass is nothing more than the semiconductor counterpart of the classical Clausius-Mossotti formula well-known in electromagnetism [67].

To illustrate the application of these ideas, next we discuss the design of a zero-gap semiconductor superlattice with linearly dispersing bands. This design is motivated by the fact that heuristically one may expect that by combining a material with a positive band gap (e.g. $Hg_{0.75}Cd_{0.25}Te$) with a material with a negative band gap (e.g. HgTe) it may be possible to design a material with a zero gap, such that the electrons experience a zero effective mass and a ultrahigh mobility.

It can be theoretically shown that the condition to have a zero gap is $M_{ef}\left(E = V_{ef}\right) = 0$ [67]. Using the analytical formulas (48)-(49), and assuming that the dispersive mass of the relevant materials is described by the linear mass approximation (43), it can be shown that the zero-gap regime occurs when the volume fraction of the HgTe circular scattering centers satisfies [67]:

$$f_{V0} = \frac{E_{v,h} + E_{v,i} - 2V_h}{E_{v,h} - E_{v,i} - 2\left(V_h - V_i\right)}. \tag{50}$$

For the particular design considered here the critical volume fraction is $f_{V0} = 0.247$. Figure 5a shows the electronic band structures of three different semiconductor superlattices characterized by certain $f_V / f_{V0}$ indicated in the insets. The superlattice period is $a = 12 a_s$, being $a_s = 0.65 nm$ the atomic lattice constant. The solid lines were obtained using the analytical effective parameters (48)-(49), whereas the discrete points were found by calculating the energy spectrum of the periodic microscopic Hamiltonian [Eq. (5)] with the plane wave method. The effective medium formulas predict almost exactly the electronic stationary states for $ka < 1.0$, and only fail near the edges of the Brillouin zone (not shown). Moreover, consistent with the previous discussion, when $f_V / f_{V0} = 1$ the band gap that separates the conduction and valence bands closes and the energy dispersion is linear. The edge of the conduction band (with $S$-type symmetry) corresponds to the energy level wherein $E = V_{ef}$, and is marked with a dashed horizontal line in the figures. Thus, for structures with $f_V / f_{V0} < 1$ (e.g., $f_V / f_{V0} = 0.9$) the superlattice is characterized by a regular band structure, whereas for $f_V / f_{V0} > 1$ (e.g., $f_V / f_{V0} = 1.1$) it is characterized by an inverted band structure. The topological transition wherein the conduction and valence bands interchange positions takes place at $f_V / f_{V0} = 1$.

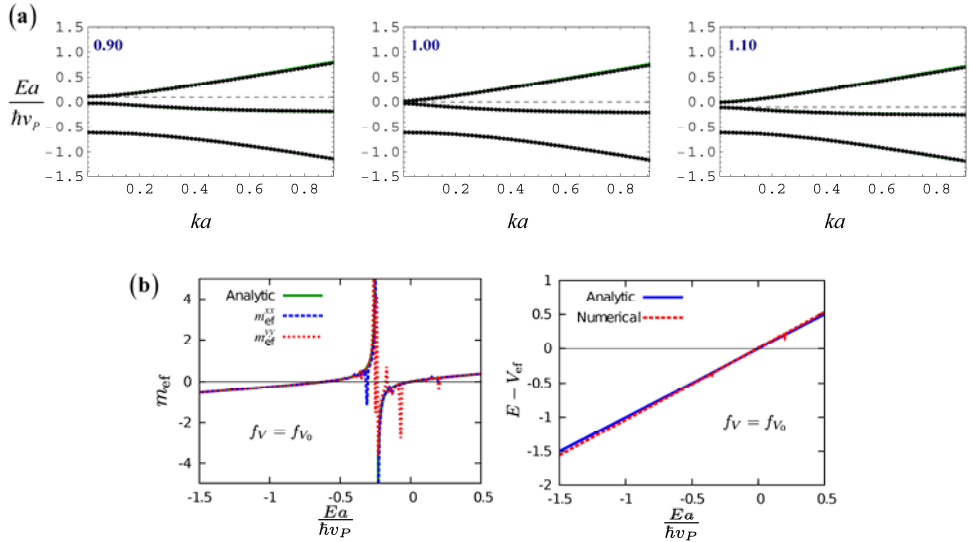

Fig. 5. (a) Electronic band structures of $Hg_{0.75}Cd_{0.25}Te$-$HgTe$ superlattices for different values of the normalized volume fraction $f_V / f_{V0}$ of the $HgTe$ inclusions shown in the insets. Solid lines: effective medium theory; Discrete points: "exact" band structure obtained with the microscopic Hamiltonian. The $HgTe$ scattering centers have circular shape and are arranged in a triangular lattice with period $a = 12 a_s$. The dashed horizontal grid line represents the edge of the hybridized conduction (*S*-type) band. (b) Effective parameters of the superlattice with $f_V = f_{V0}$ as a function of the electron energy. Adapted from Refs. [66, 67].

Figure 5b compares the analytic effective parameters [Eqs. (48)-(49)] with the effective parameters determined based on the rigorous numerical calculation of the Taylor expansion of $\mathcal{H}_{ef}$ [Eq. (45)]. Apart, from some spurious resonances in the dispersive mass the numerically calculated $V_{ef}(E)$ and $M_{ef}(E)$ agree extremely well with the analytical formulas. The spurious resonances are related to the excitation of heavy-hole type states, which are not predicted by the effective medium theory because they are dark states [66].

The zero-gap regime can have important consequences in the electron transport. In particular, due to the zero-mass property the superlattice may have a giant nonlinear response when excited by an external electromagnetic field [67]. The nonlinear effects are particularly strong up to terahertz frequencies and when the chemical potential is near the tip of the Dirac-type point [67].

### 4.4. *Homogenization of graphene superlattices*

The electron transport in graphene is determined by the energy dispersion near the Dirac points $K$ and $K'$ and is intrinsically isotropic: the electron velocity is $v_F \approx 10^6 m/s$ independent of the direction of propagation. Importantly, it has been theoretically predicted that a superlattice created by a 1D-electrostatic periodic potential (Figure 1b) can be used to collimate an electron beam with virtually no spatial spreading or diffraction [39-41], such that the electrons can only propagate along a preferred direction of space. This remarkable phenomenon can be conveniently modeled using an effective medium approach [7].

The effective Hamiltonian that describes the propagation of Dirac fermions associated with the $K$ point can be found by applying the general ideas described in section 3.4 to the two-dimensional massless Dirac Hamiltonian (6), being the microscopic electric potential $V = V(x)$ a periodic function of $x$. Note that the Dirac Hamiltonian (6) already provides an effective medium description of the electron propagation in graphene, but here we consider a second level of homogenization such that the superlattice itself can be regarded as a continuum. Because the Dirac fermions are described by a pseudo-spinor, the effective Hamiltonian $\mathcal{H}_{ef}(\mathbf{k}, E)$ is a 2×2 matrix. The computation of $\mathcal{H}_{ef}(\mathbf{k}, E)$ involves solving Eq. (25) for two independent initial macroscopic states ($S$=2). In general, this requires the discretization of the relevant partial differential equation using finite differences [7].

Similar to sections 4.1 and 4.3, to obtain some local effective parameters it is useful to expand the effective Hamiltonian in a Taylor series:

$$\mathcal{H}_{ef}(\mathbf{k}, E) \approx \mathcal{H}_{ef}(0, E) + k_x \frac{\partial \mathcal{H}_{ef}}{\partial k_x}(0, E) + k_y \frac{\partial \mathcal{H}_{ef}}{\partial k_y}(0, E). \tag{51}$$

It was found in Ref. [7] that for 1D graphene superlattices the Taylor expansion can be rewritten as $\mathcal{H}_{ef}(\mathbf{k}, E) \approx \hbar v_F \mathbf{\sigma}_{ef}(E) \cdot \mathbf{k} + V_{ef}(E)$ where $V_{ef}(E)$ is an energy dependent effective potential (a scalar) and $\mathbf{\sigma}_{ef}$ is the tensor [7]:

$$\mathbf{\sigma}_{ef} \approx v_{xx} \mathbf{\sigma}_x \hat{\mathbf{x}} + v_{yy} \mathbf{\sigma}_y \hat{\mathbf{y}}. \tag{52}$$

Here $v_{ii}$ are some scalars weakly dependent on $E$ and $\mathbf{\sigma}_i$ are the Pauli matrices. Moreover, a detailed analysis shows that

$$V_{ef}(E) \approx V_{av} - \alpha E, \quad \text{with } \alpha = v_{xx} - 1, \tag{53}$$

where $V_{av}$ is the mean value of the microscopic potential $V = V(x)$ [7].

From these results it follows that the time evolution of the macroscopic wave function in the spatial domain is determined by a modified Dirac equation [68]:

$$\left[ -i\hbar v_F \boldsymbol{\sigma}_x \frac{\partial}{\partial x} - i\hbar v_F \chi \boldsymbol{\sigma}_y \frac{\partial}{\partial y} + V_{av} \right] \cdot \Psi = i\hbar \frac{\partial}{\partial t} \Psi \;, \quad t > 0 \;, \tag{54}$$

where $\chi = v_{yy}/v_{xx}$ is by definition the *anisotropy ratio* of the superlattice. Thus, the superlattice can be regarded as a continuum described by the effective parameters $\chi$ and $V_{av}$. Comparing Eq. (54) with the original Dirac equation (6), it is seen that the main effect of the fluctuating electrostatic potential is to tailor the electron velocity in the direction perpendicular to the stratification, so that it becomes $v_F |\chi|$. Thus, the anisotropy ratio controls the velocity of the electrons along the *y*-direction. The velocity of the electrons along the *x*-direction is unaffected by the fluctuating potential due to the Klein tunneling effect [7].

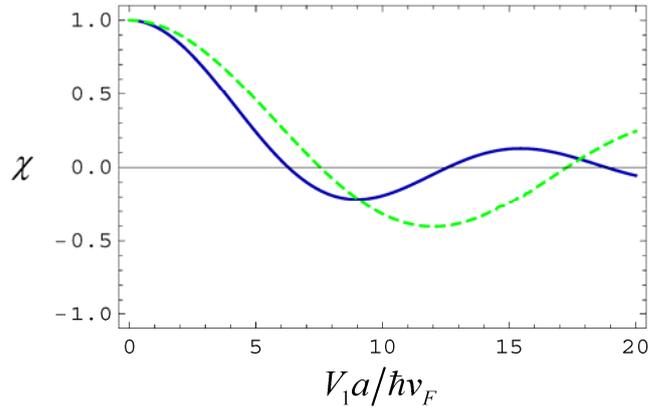

Fig. 6. Anisotropy ratio as a function of the amplitude of the oscillating part of the microscopic potential for (i) (solid line) a graphene superlattice with electric potential that alternates between the values $V_2 = -V_1$ in regions with thickness $d_1 = d_2 = a/2$. (ii) (dashed line) a sinusoidal-type electric potential with $V = V_1 \sin(2\pi x/a)$. The period of the superlattice is denoted by $a$. Adapted from Ref. [7].

Figure 6 depicts the anisotropy ratio $\chi$ as a function of the amplitude of the oscillating part of the microscopic potential, $V_{osc} = V_1$ (see Figure 1b). Two cases are considered: (i) a Krönig-Penney type potential formed by square barriers and (ii) a sinusoidal electrostatic potential. In both cases, for certain values of $V_{osc}$ the anisotropy ratio can vanish, and in these conditions the superlattice is characterized by an extreme anisotropy such that the propagation along the *y*-direction is forbidden.

The effective medium formulation (54) predicts that the energy stationary states have the dispersion [7, 68]:

$$\left| E - V_{av} \right| = \hbar v_F \sqrt{k_x^2 + \chi^2 k_y^2} \ . \tag{55}$$

Figure 7a depicts the exact energy dispersion of a graphene superlattice with a Krönig-Penney type microscopic potential with $V_1 a / \hbar v_F = 6.0$. As seen, consistent with the fact that for $V_1 a / \hbar v_F = 6.0$ the anisotropy ratio is near zero, the graphene superlattice is strongly anisotropic and the usual Dirac cone of pristine graphene is stretched along the *y*-direction. The exact energy dispersion is compared with the effective medium result (55) in Figure 7b, revealing an excellent agreement between the two formalisms.

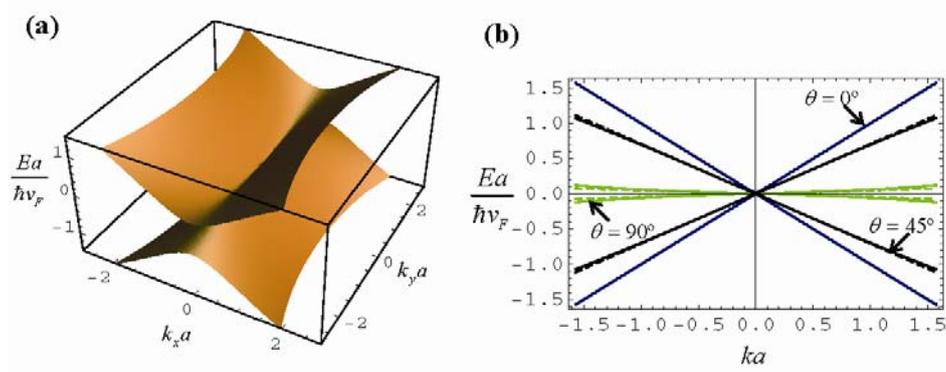

Fig. 7. (a) Exact energy dispersion of a graphene superlattice formed by square barriers with $d_1 = d_2 = a / 2$, $V_2 = -V_1$ and $V_1 a / \hbar v_F = 6.0$ (b) Dispersion of the energy eigenstates for $\mathbf{k} = k \left( \cos \theta, \sin \theta \right)$ calculated with (*i*) (solid curves) the "exact" microscopic theory. (*ii*) (dashed curves) the effective medium model based on the parameters $V_{av}$ and $\chi$. Reprinted with permission [7].

As discussed in section 3, the effective Hamiltonian describes the time evolution of macroscopic electronic states. To illustrate this property, the time evolution of an initial state was numerically determined using a finite-difference time-domain method using both the microscopic Hamiltonian (6) and the effective medium approach [Eq. (54)]. The superlattice has a sinusoidal-type profile with period *a*=10nm. The mean electric potential is set equal to zero ($V_{av} = 0$). The initial electronic state (the bright spot in Figure 8a) is taken as a Gaussian wave packet with radial width $R_G = 2.82a$. The quasi-momentum of the wave packet is set so that it propagates along the *x*-direction with the quasi-energy $E_0 a / \hbar v_F \approx 1.9$ [69].

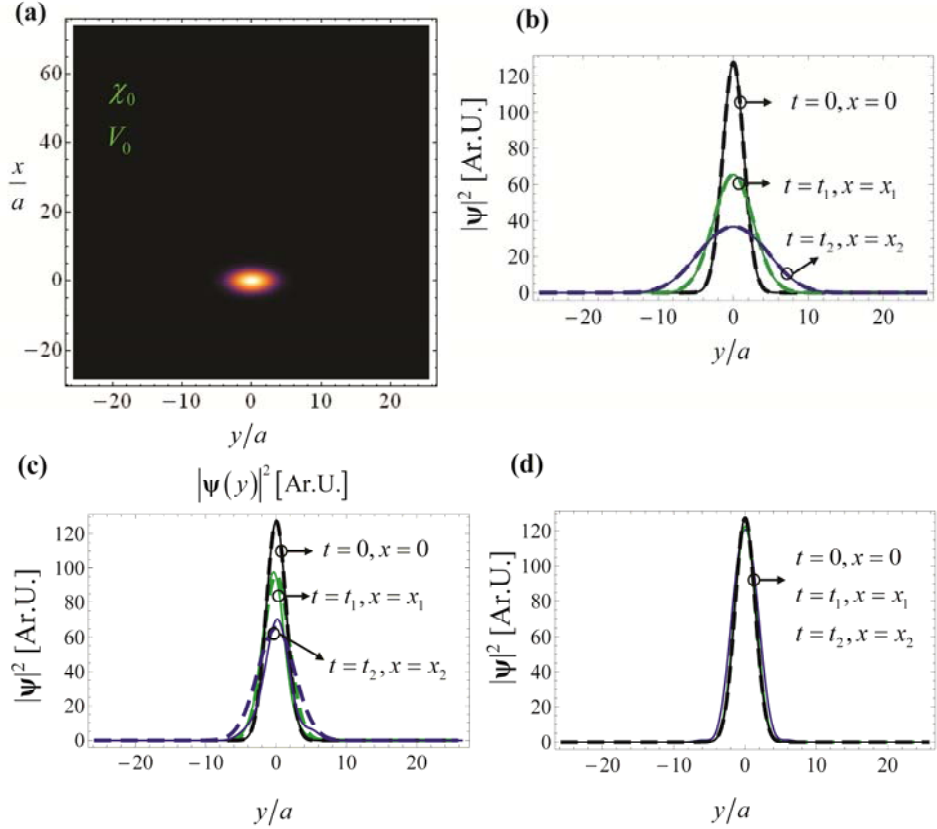

Fig. 8. (a) Sketch of the initial macroscopic electronic state (the bright spot). (b), (c) and (d) Transverse profile of the probability density function (normalized to arbitrary units) at $x = 0$ and sampled at the time instant $t = 0$, at $x_1 = 12.35a$ and $t = t_1 = 2000\Delta_t$, and at $x_3 = 24.73a$ and $t = t_2 = 4000\Delta_t$, being $\Delta_t = 0.62\,\mathrm{as}$. Panel (b) is for pristine graphene ($\chi = 1$, $V = 0$), panel (c) is for a superlattice with $\chi = 0.7$ (in the microscopic model $V_{osc}a/\hbar v_F \approx 3.58$), and panel (d) is for a superlattice with $\chi = 0$ (in the microscopic model $V_{osc}a/\hbar v_F \approx 7.55$). In all the plots, the dashed lines represent the microscopic theory results, and the solid thick lines represent the effective medium results. The microscopic potential has a sinusoidal profile with period $a = 10\,nm$. Reprinted with permission [69].

Figures 8b, 8c and 8d show the calculated transverse profiles of the probability density function sampled at different instants of time, for pristine graphene ($\chi = 1$) and for superlattices with anisotropy ratio $\chi = 0.7$ and $\chi = 0$, respectively. As expected, for pristine graphene (Figure 8b) the time evolution of the initial electronic state causes the wave packet to diffract and increase its characteristic size. In pristine graphene the group velocity is independent of the direction of propagation, and hence there is no preferred direction of motion.

Quite differently, in a graphene superlattice with $\chi = 0$ (Figure 8d) the electronic state is unaffected by diffraction and the shape of the wave front does not change with time. Crucially, the time evolution predicted by the exact microscopic Hamiltonian (dashed lines) is practically coincident with that predicted by the effective medium formulation (solid lines). Indeed, as further demonstrated in the example of Figure 9a, the effective medium Hamiltonian can determine very accurately the time evolution of an initial electronic state that is less localized than the period $a$. When the characteristic size of the initial state is comparable or less than $a$, the effective medium theory and the microscopic theory give diverging results (Figures 9b and 9c).

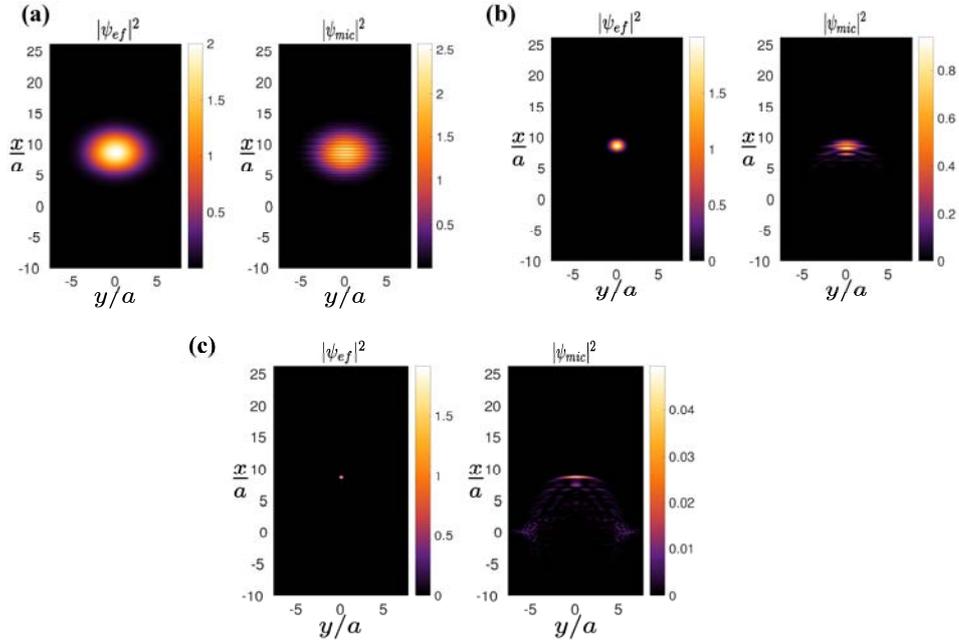

Fig. 9. Profiles of the probability density function after the initial state (initially centered at the origin) propagates during $1000\Delta_t$ seconds. Left: effective medium result. Right: microscopic theory result; The superlattice is characterized by an extreme anisotropy $\chi = 0$ (in the microscopic model $V_{osc}a/\hbar v_F \approx 7.55$) and has the same parameters as in Fig. 8. The initial state has the radial width (a) $R_G = 4a$ (b) $R_G = a$ and (c) $R_G = 0.25a$. Adapted from Ref. [69].

Importantly, the effective medium formulation can be extended to graphene heterostructures formed by non-uniform (with parameters varying in space) superlattices, as well as to characterize the scattering of electron waves by graphene nanostructures. Detailed examples are reported in Refs. [68-69].

## 5. Homogenization near the corners of the Brillouin zone

In section 4 the "low energy" electronic states are always associated with a wave vector near the origin ($\Gamma$ point, $\mathbf{k} = 0$). This property also applies to the graphene superlattice example. Indeed, within the validity of the microscopic Hamiltonian (6), the wave vector $\mathbf{k}$ is measured with respect to the $K$ point of the Brillouin zone, and hence it is near zero in the vicinity of the Dirac point. Importantly, at a more fundamental level one should take into account that graphene is itself a two-dimensional periodic material described by the Schrödinger equation, with an electric potential $V$ that has the honeycomb symmetry with two carbon atoms per unit cell [35]. Within this more fundamental description, the physics of graphene is determined by the $K$ and $K'$ points at the edges of the Brillouin zone, which are the points where the conduction and valence bands of graphene meet [35].

The objective of this Section is to discuss how to homogenize systems wherein the relevant wave phenomena are determined by points of the $\mathbf{k}$-space different from the origin. To better illustrate the ideas, the discussion is focused on the problem of homogenization of "artificial graphene": a two-dimensional electron gas (2DEG) nanopatterned with scattering centers organized in a lattice with the honeycomb symmetry (Figure 1c) [33]. Such a superlattice has an electronic band structure analogous to graphene, and exhibits linearly dispersing bands near the $K$ and $K'$ points at some Dirac energy, $E_D$ [33]. The electron wave propagation in artificial graphene may be described by the Hamiltonian (5), being $m$ the effective electron mass in the electron gas (which is assumed independent of the position) and $V$ is a periodic function taken to be $V_0$ inside the circular scattering centers, and zero outside [33, 66]. The circular scattering centers have radius $R$ and the nearest neighbor distance is $a$.

The effective Hamiltonian $\mathcal{H}_{ef}(\mathbf{k}, E)$ of the modulated 2DEG can be determined in a straightforward manner following the ideas of section 3.3. The calculation of $\mathcal{H}_{ef}(\mathbf{k}, E)$ requires the use of numerical methods [66]. It should be noted that $\mathcal{H}_{ef}(\mathbf{k}, E)$ is a scalar function because the wave function in Eq. (5) is a scalar. Moreover, since the relevant physics is determined by the corners of the first Brillouin zone, the set B.Z. in Eq. (11) should be taken equal to a translated version of the first Brillouin such that both the $K$ and $K'$ points are interior to the set.

In order to describe the wave dynamics of the low energy states near the $K$ point, it is tempting to mimic the ideas of section 4 and expand $\mathcal{H}_{ef}(\mathbf{k}, E)$ in a Taylor series around $\mathbf{k} = K$. Unfortunately, such an approach does not work. Indeed, it turns out that the effective Hamiltonian $\mathcal{H}_{ef}(\mathbf{k}, E)$ has a direction-dependent resonant behavior near the $K$ point and therefore the Taylor series is useless. This property is illustrated in Figure 10, which depicts $\mathcal{H}_{ef}(\mathbf{k}, E) - E$ as a function of $\mathbf{q}$, being $\mathbf{q} = \mathbf{k} - K$ the wave vector measured with respect to the $K$ point. Thus, it follows that the system is strongly spatially dispersive at the corners of the Brillouin zone.

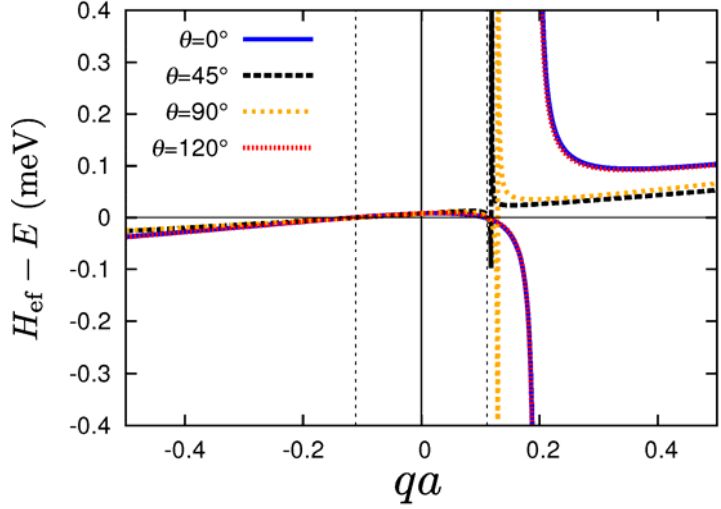

Fig. 10. Effective Hamiltonian near the $K$ point as a function of the normalized wave vector $\mathbf{q} = q(\cos\theta, \sin\theta)$ for different directions of propagation and $E = -0.33\,meV$, $V_0 = -0.8\,meV$, and $R/a = 0.35$. The vertical dashed lines indicate the zeros of $\mathcal{H}_{ef} - E$. Reprinted with permission [66].

Remarkably, it is possible to avoid the strong spatial dispersion effects by considering a pseudo-spinor description of the wave propagation, such that the macroscopic wave function becomes a two-component vector [66]. This can be done with a generalization of the theory of section 3, as described next.

The strong spatial dispersion of $\mathcal{H}_{ef}(\mathbf{k}, E)$ can be attributed to the fact that the wave function can have significant fluctuations within each unit cell because there are two inequivalent scattering sites per cell. This property suggests that the definition of macroscopic state may be too restrictive for the system under study. Indeed, a macroscopic electronic state cannot be more localized than the lattice period, and hence the two sub-lattices of the modulated 2DEG are not distinguished in the effective medium theory. Inspired by the pseudo-spinor formalism of graphene [35], next we introduce generalized macroscopic states that allow for the discrimination of the two sub-lattices.

As illustrated in Figure 11, the idea is to consider a partition of the unit cell into two regions ($i$=1,2). Each region is described by a characteristic periodic function, $\chi_i(\mathbf{r})$, such that $\chi_1 + \chi_2 = 1$. The function $\chi_i$ assumes the value 1 in the $i$-th region and is zero otherwise. By definition a *generalized macroscopic state* is of the form $\psi = \psi_1\chi_1 + \psi_2\chi_2$, with $\psi_i = \{\psi_i\}_{av}$ ($i$=1,2). Evidently, this definition extends that of section 3, and allows the macroscopic states to be as localized as each individual scattering center.

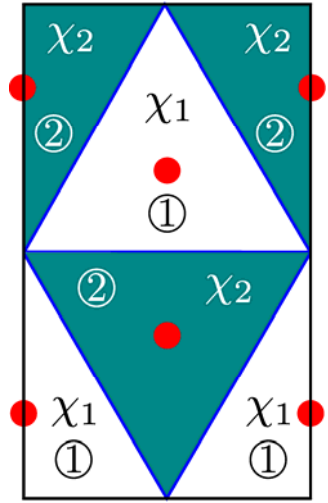

Fig. 11. Partition of space determined by the characteristic functions $\chi_1$ and $\chi_2$. The circular dots represent the scattering centers of the honeycomb lattice. Reprinted with permission [66].

To obtain an effective medium description of the generalized macroscopic states, the Schrödinger equation (1) is rewritten as [66]:

$$\underbrace{\begin{pmatrix} \chi_1\hat{H}\chi_1 & \chi_1\hat{H}\chi_2 \\ \chi_2\hat{H}\chi_1 & \chi_2\hat{H}\chi_2 \end{pmatrix}}_{\hat{H}_g} \cdot \psi_g = i\hbar\frac{\partial}{\partial t}\psi_g, \qquad \psi_g = \begin{pmatrix} \chi_1\psi \\ \chi_2\psi \end{pmatrix}, \qquad (56)$$

where $\hat{H}_g$ is a generalized two-component microscopic Hamiltonian. The two-component wave function $\psi_g$ may be regarded as a microscopic pseudo-spinor. Each of the components of $\psi_g$ is associated with a specific sub-lattice of the modulated 2DEG.

Similar to section 3, the goal is to introduce a generalized effective Hamiltonian $\hat{H}_{g,ef}$ that describes the time evolution of generalized macroscopic initial states $\psi_{t=0} = \psi_{1,t=0}\chi_1 + \psi_{2,t=0}\chi_2$. The effective Hamiltonian must ensure that $\left(\hat{H}_{g,ef}\cdot\Psi_g\right)(\mathbf{r},t)=\left\{\left(\left(\hat{H}_g\cdot\psi_g\right)(\mathbf{r},t)\right)\right\}_{\text{av}}$ for any initial macroscopic state (Figure 2). Here, $\Psi_g=\left\{\psi_g\right\}_{\text{av}}$ is the macroscopic pseudo-spinor.

To determine the effective Hamiltonian in the Fourier domain, $\boldsymbol{\mathcal{H}}_{g,ef}\left(\mathbf{k},E\right)$, it is supposed that $-i\hbar\psi_{i,t=0}=f_i e^{i\mathbf{k}\cdot\mathbf{r}}$ where the weighting factors, $f_i$, can be chosen arbitrarily. Then, calculating the unilateral Fourier (Laplace) transform in time of Eq. (56) it is found that:

$$\left(\hat{H}_g-E\right)\cdot\psi_g=\begin{pmatrix}\chi_1 f_1\\\chi_2 f_2\end{pmatrix}e^{i\mathbf{k}\cdot\mathbf{r}}. \tag{57}$$

Because the effective Hamiltonian must guarantee that $\left(\hat{H}_g\cdot\psi_g\right)_{\text{av}}=\boldsymbol{\mathcal{H}}_{g,ef}\left(\mathbf{k},E\right)\cdot\psi_{g,\text{av}}$, the above equation implies that [66]:

$$\left[\boldsymbol{\mathcal{H}}_{g,ef}\left(\mathbf{k},E\right)-E\right]\cdot\psi_{g,\text{av}}=\frac{1}{2}\begin{pmatrix}f_1\\f_2\end{pmatrix}, \tag{58}$$

where $\psi_{g,\text{av}}$ and $\left(\hat{H}_g\cdot\psi_g\right)_{\text{av}}$ are given by formulas analogous to Eq. (24). Thus, by solving the microscopic problem (57) for two independent sets of weighting factors $f_i$ and by finding the corresponding $\psi_{g,\text{av}}$ using Eq. (24a), it is possible to compute the two-component effective Hamiltonian with Eq. (58). These ideas are developed in detail in Ref. [66].

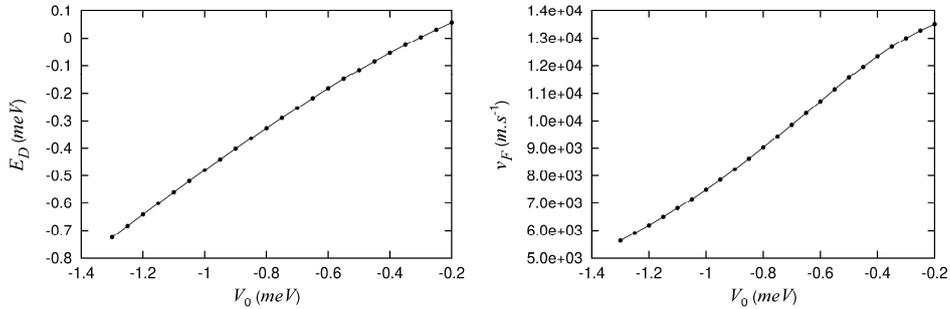

Fig. 12. Dirac energy (left) and Fermi velocity (right) as a function of the potential $V_0$ calculated using the effective medium theory for the Dirac cone near the $K$ point. The radius of the scattering centers is $R/a=0.35$. Adapted from Ref. [66].

Importantly, in contrast with the single component Hamiltonian (Figure 10), the two-component Hamiltonian $\boldsymbol{\mathcal{H}}_{g,ef}\left(\mathbf{k},E\right)$ is a smooth function of the wave vector near the Dirac points $K$ and $K'$, and thus it can be approximated by a first order Taylor series around the Dirac points. It was demonstrated in Ref. [66] that such an expansion leads –after a suitable renormalization of the pseudo-spinor– to the same massless two-dimensional Dirac equation (6) that describes the physics of graphene. Hence, the generalized effective medium theory described here gives a "first principles" demonstration that the low-energy electron wave propagation in "artificial graphene" is described by the Dirac massless equation. Moreover, from the coefficients of the Taylor expansion of $\mathcal{H}_{g,ef}(\mathbf{k}, E)$ it is possible to compute the dependence of the equivalent Fermi velocity $v_F$ and of the Dirac energy $E_D$ as a function of the height of the potential wells $-V_0$ associated with the scattering centers [66]. An illustrative example is reported in Figure 12.

## 6. Quadratic forms

In the following we discuss how to determine some quadratic forms of the electromagnetic fields (in case of light waves) or of the wave function (in case of quantum systems) relying on the effective medium framework. Examples of such quadratic forms are the electromagnetic energy density or the probability density function.

### 6.1. *Hermitian symmetry*

A key property of the Hamiltonian operator $\hat{H}$ in quantum physics is that it is Hermitian. The Hermitian symmetry guarantees that the wave function $\psi(t)$ differs from the initial state $\psi(0)$ by a unitary transformation, and hence the norm $\langle \psi | \psi \rangle$ is independent of time. In particular, $\left| \psi(\mathbf{r}, t) \right|^2$ is the probability density function for the particle position, and its integral over the space coordinates is independent of time.

The effective Hamiltonian $\hat{H}_{ef}$ defined in section 3 describes the time dynamics of the envelope-function $\Psi(\mathbf{r}, t)$. As previously mentioned, in contrast with the microscopic Hamiltonian, the action of the effective Hamiltonian on the macroscopic state vector is not "local" in time and depends on the past history of the state vector. Thus, the operator $\hat{H}_{ef}$ has a completely different nature than the microscopic Hamiltonian, and the notion of Hermitian symmetry does not directly apply. Yet, in the Fourier domain the Hermitian symmetry is recovered, specifically for quantum systems $\mathcal{H}_{ef}(\mathbf{k}, E)$ is Hermitian symmetric.

To prove such a property, we rewrite Eq. (25) as:

$$\left( \hat{H} - E \right) \cdot \boldsymbol{\psi} = \mathbf{1}_{S \times S} \, e^{i\mathbf{k} \cdot \mathbf{r}}, \tag{59}$$

where $\mathbf{1}_{S \times S}$ is the identity matrix of rank $S$ and $\boldsymbol{\psi} = \begin{bmatrix} \psi^{(1)} & \dots & \psi^{(S)} \end{bmatrix}$ should be understood as a $S \times S$ matrix whose columns are the solutions $\psi^{(l)}$ of Eq. (25)

for $-i\hbar\psi^{(l)}_{t=0}=e^{i\mathbf{k}\cdot\mathbf{r}}\mathbf{u}_l$. Multiplying both sides of Eq. (59) by $\boldsymbol{\psi}^{\dagger}$ and integrating the resulting expression over the unit cell $\Omega$ it is found that:

$$\boldsymbol{\psi}^{\dagger}_{\mathrm{av}}=\left[\left\langle\psi^{(i)}\mid\hat{H}-E\mid\psi^{(j)}\right\rangle\right]_{i,j}. \tag{60}$$

The right-hand side represents a $S\times S$ matrix with generic $i,j$ element given by $\left\langle\psi^{(i)}\mid\hat{H}-E\mid\psi^{(j)}\right\rangle$, where

$$\left\langle f\mid g\right\rangle=\frac{1}{V_{cell}}\int_{\Omega}d^{N}\mathbf{r}\,f^{*}\left(\mathbf{r}\right)g\left(\mathbf{r}\right) \tag{61}$$

is the canonical inner product in the unit cell. The matrix $\boldsymbol{\psi}_{\mathrm{av}}$ is given by $\boldsymbol{\psi}_{\mathrm{av}}=\left[\psi^{(1)}_{\mathrm{av}}\quad...\quad\psi^{(S)}_{\mathrm{av}}\right]$ where the column vectors are defined as in section 3.4.

For quantum systems the operator $\hat{H}$ is Hermitian with respect to the canonical inner product, and hence the matrix on the right-hand side of Eq. (60) is Hermitian. This property shows that $\boldsymbol{\psi}_{\mathrm{av}}$ is also a Hermitian matrix, and from Eq. (26) it finally follows that $\mathcal{H}_{ef}\left(\mathbf{k},E\right)$ is Hermitian symmetric, as we wanted to prove. It is important to highlight that the Hermitian symmetry does not apply to the electromagnetic case, because the operator $\hat{H}$ for light waves [Eq. (4)] is not Hermitian with respect to the canonical inner product.

## 6.2. *The probability density function and the probability current for quantum systems*

A crucial observation is that the operation of spatial averaging does not commute with the multiplication operation. Thus, in general the averaged probability density function calculated with the microscopic theory cannot be identified with the squared amplitude of the envelope-function $\Psi\left(\mathbf{r},t\right)=\left\{\psi\left(\mathbf{r},t\right)\right\}_{\mathrm{av}}$:

$$\Psi^{*}\cdot\Psi\left(\mathbf{r},t\right)\neq\left\{\psi^{*}\cdot\psi\left(\mathbf{r},t\right)\right\}_{\mathrm{av}}. \tag{62}$$

In other words, typically $\left|\Psi\left(\mathbf{r},t\right)\right|^{2}$ cannot be regarded as the probability density function for the particle position and thus its integral over the spatial coordinates may be time dependent [7].

This result may seem at first surprising, but actually it is a consequence of the fact that $\Psi\left(\mathbf{r},t\right)$ only describes the envelope of the microscopic wave function. Even though when the initial state is macroscopic one has $\psi=\Psi$ at the time origin $t=0$, as time passes $\psi\left(\mathbf{r},t\right)$ may depart significantly from a pure macroscopic state and hence it can have strong fluctuations on the scale of the unit cell. In contrast, $\Psi(\mathbf{r},t)$ varies always slowly on the scale of the unit cell and this explains the property (62).

Ultimately, Eq. (62) is a consequence of the dispersive nature of the effective Hamiltonian, which usually is energy dependent. If $\mathcal{H}_{ef}(\mathbf{k},E)$ is weakly dependent on $E$, then from the analysis of section 6.1 one may conclude that $\hat{H}_{ef}$ is also Hermitian, and in that case $\left|\Psi(\mathbf{r},t)\right|^2$ can be identified with the probability density function.

Importantly, for the energy stationary states the averaged probability density function is precisely determined by the effective Hamiltonian. Specifically, it can be proven that for Bloch stationary states ($\psi$) of the microscopic Hamiltonian one has the following exact result (see Ref. [66] and the supplementary materials of Ref. [68]):

$$\left\{\psi^*\cdot\psi\right\}_{\mathrm{av}}=\Psi^*\cdot\left(1-\frac{\partial\mathcal{H}_{ef}}{\partial E}\right)\cdot\Psi\,. \tag{63}$$

Thus, for Bloch states the averaged probability density function can be written in terms of the macroscopic wave function and of the energy derivative of the effective Hamiltonian. In the absence of energy dispersion one obtains $\left\{\psi^*\cdot\psi\right\}_{\mathrm{av}}=\Psi^*\cdot\Psi$, consistent with the previous discussion.

Moreover, supposing that the microscopic Hamiltonian is of the form $\hat{H}=\hat{H}\left(-i\nabla,\mathbf{r}\right)$ it is possible to demonstrate that for Bloch stationary states the spatially averaged probability current $\mathbf{j}=\psi^*\cdot\hbar^{-1}\partial\hat{H}/\partial\mathbf{k}\cdot\psi$ with $\mathbf{k}=-i\nabla$ exactly satisfies:

$$\left\{\psi^*\cdot\frac{1}{\hbar}\frac{\partial\hat{H}}{\partial\mathbf{k}}\left(-i\nabla,\mathbf{r}\right)\cdot\psi\right\}_{\mathrm{av}}=\Psi^*\cdot\frac{1}{\hbar}\frac{\partial\mathcal{H}_{ef}}{\partial\mathbf{k}}\cdot\Psi\,. \tag{64}$$

Hence, in the macroscopic framework the spatially averaged probability current can be identified with the right-hand side of Eq. (64). Applications of Eqs. (63)-(64) are discussed in Refs. [66, 68, 70].

In summary, for Bloch stationary states both the probability density function and the probability current are rigorously determined by the effective medium theory. Note that for Bloch waves one has:

$$\left\{\psi^*\cdot\psi\right\}_{\mathrm{av}}=\frac{1}{V_{cell}}\int_{\Omega}\psi^*\cdot\psi\,d^N\mathbf{r}\,. \tag{65}$$

A similar expression can be written for $\left\{\psi^*\cdot\dfrac{1}{\hbar}\dfrac{\partial\hat{H}}{\partial\mathbf{k}}\left(-i\nabla,\mathbf{r}\right)\cdot\psi\right\}_{\mathrm{av}}$.

### 6.3. *The energy density and the Poynting vector for electromagnetic systems*

The light counterparts of the probability density function and of the probability current are the electromagnetic energy density and the Poynting vector. Interestingly, similar to the case of electron waves, such quadratic forms of the fields can be rigorously determined in a time-harmonic stationary regime. Specifically, let us consider a generic periodic electromagnetic metamaterial that is described at the microscopic level by a lossless permittivity function $\varepsilon = \varepsilon(\mathbf{r}, \omega)$ and by the permeability $\mu = \mu_0$. As discussed in section 3.5, in the effective medium approach such systems are fully characterized by a nonlocal dielectric function $\varepsilon_{ef}(\mathbf{k}, \omega)$. Let us consider a generic Bloch electromagnetic mode of the metamaterial with a time-harmonic variation $e^{-i\omega t}$ and associated with the complex microscopic electromagnetic fields $\mathbf{e}$ and $\mathbf{b}$. The spatially averaged electromagnetic energy density and Poynting vector are defined as follows:

$$W_{av} = \frac{1}{4V_{cell}} \int_\Omega \frac{|\mathbf{b}|^2}{\mu_0} d^3\mathbf{r} + \frac{1}{4V_{cell}} \int_\Omega \frac{\partial}{\partial\omega}(\omega\varepsilon)|\mathbf{e}|^2 d^3\mathbf{r}, \qquad (66)$$

$$\mathbf{S}_{av} = \frac{1}{V_{cell}} \int_\Omega \frac{1}{2} \mathrm{Re}\left\{\mathbf{e} \times \frac{\mathbf{b}^*}{\mu_0}\right\} d^3\mathbf{r}. \qquad (67)$$

Then, in analogy with section 6.2, it is possible to prove that $W_{av}$ and $\mathbf{S}_{av}$ can be written in terms of the macroscopic fields and of the effective dielectric function as follows [71-73]:

$$W_{av} = \frac{1}{4} \frac{|\mathbf{B}|^2}{\mu_0} + \frac{1}{4} \mathbf{E}^* \cdot \frac{\partial}{\partial\omega}(\omega\overline{\varepsilon}_{ef}) \cdot \mathbf{E}, \qquad (68)$$

$$\mathbf{S}_{av} \cdot \hat{l} = \frac{1}{2} \mathrm{Re}\left\{\mathbf{E} \times \frac{\mathbf{B}^*}{\mu_0}\right\} \cdot \hat{l} - \frac{1}{4}\omega\mathbf{E}^* \cdot \frac{\partial\overline{\varepsilon}_{ef}}{\partial k_l}(\mathbf{k}, \omega) \cdot \mathbf{E}, \quad l = x, y, z. \qquad (69)$$

In the above, $\mathbf{E} = \{\mathbf{e}\}_{av}$ and $\mathbf{B} = \{\mathbf{b}\}_{av}$ represent the macroscopic fields associated with the Bloch mode with wave vector $\mathbf{k}$. It is emphasized that Eqs. (68)-(69) are exact for lossless electromagnetic systems [71, 73].

Moreover, if the metamaterial response can be assumed to a good approximation local so that Eq. (36) holds, then Eqs. (68)-(69) reduce to:

$$W_{av} = \frac{1}{4} \frac{\partial}{\partial\omega}(\omega\mu_0\mu_L)|\mathbf{H}_L|^2 + \frac{1}{4} \frac{\partial}{\partial\omega}(\omega\varepsilon_0\varepsilon_L)|\mathbf{E}|^2, \qquad (70)$$

$$\mathbf{S}_{av} = \frac{1}{2}\text{Re}\{\mathbf{E} \times \mathbf{H}_L^*\}, \qquad (71)$$

where $\mathbf{H}_L \equiv \mu_0^{-1}\mu_L^{-1}\mathbf{B}$ is by definition the macroscopic local magnetic field. It is relevant to highlight that the formulas (68)-(69) for the stored energy density and for the Poynting vector are coincident with well known formulas for an electromagnetic spatially dispersive continuum [48]. Similarly, Eqs. (70)-(71) are coincident with the classical textbook formulas for the energy density and for the Poynting vector in dispersive isotropic dielectrics [48]. Thus, the theory of Refs. [71-73] proves that these classical textbook formulas for a continuum describe precisely the spatially averaged microscopic energy density and Poynting vector within the effective medium theory discussed here.

To illustrate the described ideas, next we consider a two-dimensional metamaterial formed by a square array of high-index dielectric cylindrical inclusions with radius $R/a = 0.435$, permittivity $\varepsilon_d = 50.47$ and permeability $\mu = 1$, embedded in a host material characterized by a lossless Drude dispersion model with $\varepsilon_h = 1 - \omega_p^2/\omega^2$. The lattice period is $a$ and the normalized plasma frequency is taken equal to $\omega_p a/c = 1.0$. The geometry of the unit cell is depicted in the inset of Fig. 13d. The magnetic field is polarized along the $z$-direction (parallel to the cylinders axes) and the electric field is in the $xoy$ plane. Similar to the example of Section 4.1, to a good approximation this metamaterial has a local response and is characterized by local effective parameters that satisfy $\varepsilon_L \approx \mu_L$ in a broad frequency range (see Figure 13a). In particular, below the plasma frequency $\omega < \omega_p$ the material behaves as double negative material with simultaneously negative permittivity and permeability [73].

Figures 13b and 13c show a comparison between the numerically calculated spatially averaged Poynting vector and energy density [Eqs. (66)-(67)] and the same quantities determined using the nonlocal electromagnetic continuum formulas [Eqs. (68)-(69)]. The nonlocal dielectric function is obtained as explained in section 3.5 and the macroscopic fields are found by spatially averaging the microscopic Bloch modes. As seen, the numerical results confirm that Eqs. (66)-(67) and Eqs. (68)-(69) give coincident results. Figures 13b and 13c also reveal that the results predicted by the local approximation [Eqs. (70)-(71)] are quite accurate, especially below the plasma frequency. In contrast, the Poynting vector computed with the erroneous formula $\mathbf{S}_{av} = 1/2\,\text{Re}\{\mathbf{E} \times \mathbf{B}^*\}$ proposed in Ref. [74] to describe the energy density flux in metamaterials gives a completely disparate result. Indeed, such a formula neglects the artificial magnetism induced by the high-permittivity cylinders, and incorrectly implies that negative refraction and backward propagation are impossible in the metamaterial [71, 72, 74].

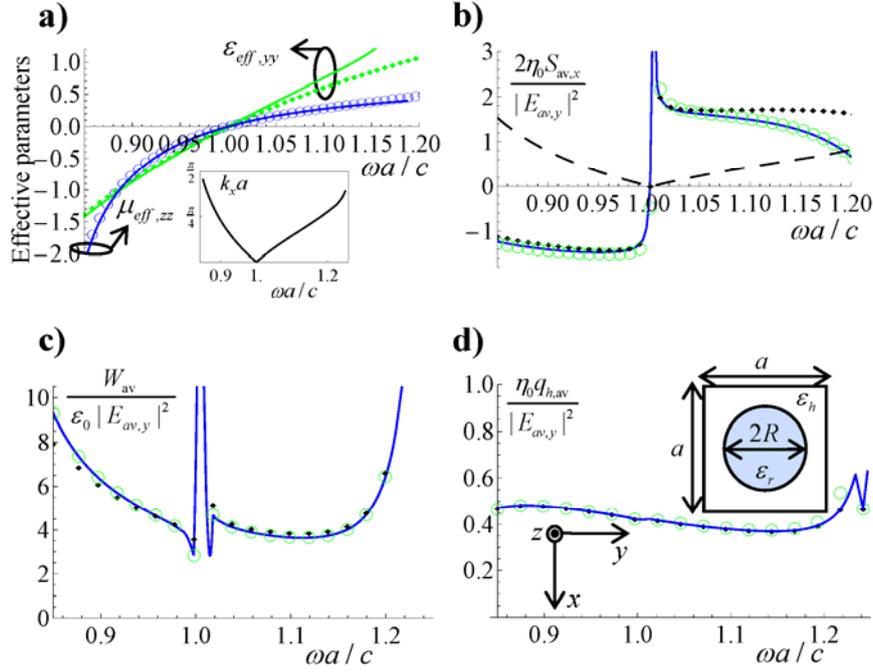

Fig. 13 (a) Local permittivity $\varepsilon_L$ and local permeability $\mu_L$ versus the normalized frequency $\omega a / c$. The discrete symbols are calculated with the effective medium theory and the solid lines are calculated using the Clausius-Mossotti formulas. The inset shows the wave vector $k_x$ as a function of frequency. (b) $x$-component of the Poynting vector calculated using: (i) Averaged microscopic Poynting vector (solid line) [Eq. (67)]; (ii) Nonlocal homogenization (circles) [Eq. (69)]; (iii) Local effective parameters (diamonds) [Eq. (71)]; (iv) Result obtained with the erroneous definition of the Poynting vector of Ref. [74] (dashed line). (c) Electromagnetic energy density calculated using: (i) Averaged microscopic energy density (solid line) [Eq. (66)]; (ii) Nonlocal homogenization (circles) [Eq. (68)]; (iii) Local effective parameters (diamonds) [Eq. (70)]. (d) Heating rate calculated using: (i) Averaged microscopic heating rate (solid line) [Eq. (72)]; (ii) Nonlocal homogenization (circles) [Eq. (73)]; (iii) Local effective parameters (diamonds) [Eq. (74)]. The unit cell geometry is shown in the inset. Reprinted with permission [73].

It is emphasized that the equivalence between Eqs. (66)-(67) and Eqs. (68)-(69) is only observed for lossless materials. In case of material loss, the equivalence is only approximate [73]. Interestingly, in the lossy case the effective medium theory can predict exactly the heating rate due to the material absorption in time-harmonic regime. Specifically, defining the spatially averaged microscopic heating rate as:

$$q_{av} = \frac{1}{V_{cell}} \int_\Omega \frac{\omega}{2} \varepsilon''(\mathbf{r}) |\mathbf{e}(\mathbf{r})|^2 \, d^3\mathbf{r} \,, \tag{72}$$

it can be shown that it is exactly coincident with the result predicted by the corresponding continuum formula [71-73]:

$$q_{av} = \frac{1}{2} \mathrm{Re} \left\{ -i\omega \mathbf{E}^* \cdot \overline{\varepsilon}_{ef} (\mathbf{k}, \omega) \cdot \mathbf{E} \right\}. \tag{73}$$

When the material response is approximately local the continuum formula becomes simply:

$$q_{av} = \frac{1}{2} \omega \varepsilon_0 \varepsilon_L'' (\omega) |\mathbf{E}|^2 + \frac{1}{2} \omega \mu_0 \mu_L'' (\omega) |\mathbf{H}_L|^2 \,. \tag{74}$$

Figure 13d shows the comparison among these three definitions (72)-(74) for a host medium with normalized damping frequency $\Gamma / \omega_p = 0.1$ and a cylinder permittivity $\varepsilon_d = 50.47 + 0.1i$. The numerical results confirm that Eqs. (72)-(73) give coincident results. Figure 13d also reveals that Eq. (74) is nearly exact for this example.

## 7. Summary

We described a completely general self-consistent approach to characterize the wave propagation in periodic systems from an effective medium perspective. The theory relies on the introduction of an effective Hamiltonian operator that regards the system as a continuum. The effective Hamiltonian describes exactly the time evolution of the wave packet envelope when the initial state is less localized than the lattice period. In addition, the effective Hamiltonian determines completely the band diagram of the time-stationary states of the periodic system. The described theory can be applied to a wide range of physical systems. Here, we illustrated its application to the homogenization of electromagnetic metamaterials and to the homogenization of semiconductor and graphene superlattices. In particular, it was highlighted that the effective medium description can be often simplified based on the extraction of local effective parameters from the effective Hamiltonian, even for systems wherein the relevant physics is determined by points at the corners of the Brillouin zone. Finally, it was shown that the effective Hamiltonian determines exactly some quadratic forms related to the energy density and to the energy transport.